\documentclass[%
superscriptaddress,
 amsmath,amssymb,
 aps,
]{revtex4-2}

\usepackage{graphicx}
\usepackage{dcolumn}
\usepackage{bm}
\usepackage{xcolor}

\begin{document}
\title{Effect of Transport Mechanism on the Coarsening of Bicontinuous Structures: A Comparison between Bulk and Surface Diffusion}
\author{W. Beck Andrews}
\affiliation{Department of Materials Science and Engineering, University of Michigan, Ann Arbor, MI 48109, USA}%
\author{Kate L.M. Elder}
\affiliation{Department of Materials Science and Engineering, Northwestern University, Evanston, IL 60208, USA}%
\author{Peter W. Voorhees}
\affiliation{Department of Materials Science and Engineering, Northwestern University, Evanston, IL 60208, USA}%
\author{Katsuyo Thornton}
\affiliation{Department of Materials Science and Engineering, University of Michigan, Ann Arbor, MI 48109, USA}%
\email{kthorn@umich.edu}

\date{\today}

\begin{abstract}
[This article was published as Phys.\ Rev.\ Materials 4, 103401 on 14 October 2020; \url{https://doi.org/10.1103/PhysRevMaterials.4.103401}]
Coarsening of bicontinuous microstructures is observed in a variety of systems, such as nanoporous metals and mixtures that have undergone spinodal decomposition.
To better understand the morphological evolution of these structures during coarsening, we compare the morphologies resulting from two different coarsening mechanisms, surface and bulk diffusion.
We perform phase-field simulations of coarsening via each mechanism in a two-phase mixture at nominal volume fractions of 50\%-50\% and 36\%-64\%, and the simulated structures are characterized in terms of topology (genus density), the interfacial shape distribution, structure factor, and autocorrelations of phase and mean curvature.
We observe self-similar evolution of morphology and topology and agreement with the expected power laws for dynamic scaling, in which the characteristic length scale increases over time proportionally to $t^{1/4}$ for surface diffusion and $t^{1/3}$ for bulk diffusion.
While we observe the expected difference in the coarsening kinetics, we find that differences in self-similar morphology due to coarsening mechanism are relatively small, although typically they are larger at 36\% volume fraction than at 50\% volume fraction.
In particular, we find that bicontinuous structures coarsened via surface diffusion have lower scaled genus densities than structures coarsened via bulk diffusion.
We also compare the self-similar morphologies to those in literature and to two model bicontinuous structures, namely, constant-mean-curvature surfaces based on the Schoen G minimal surface and random leveled-wave structures.
The average scaled mean curvatures of these model structures agree reasonably with those of the coarsened structures at both 36\% and 50\%, but we find substantial disagreements in the scaled genus densities and the standard deviations of mean curvature.

\end{abstract}

\maketitle

\section{Introduction}
Coarsening -- the spontaneous evolution of a two-phase system driven by the reduction of its total interfacial energy -- is often assumed to exhibit self-similarity at sufficiently long times after phase separation \cite{Bray2002}.
During self-similar evolution, statistical characteristics of the morphology are time-invariant when scaled by a characteristic length scale $\ell$, which evolves in time according to a power law $\ell \propto t^{1/p}$, where the exponent $p$ depends on the dynamics of coarsening \cite{Herring1950,Bray2002}.
These predictions, self-similarity and power law kinetics, are expected to hold for coarsening via a variety of possible transport mechanisms, such as surface diffusion \cite{Geslin2019}, bulk diffusion \cite{Kwon2010}, and hydrodynamic flow \cite{Henry2018}.
The power law exponent $p$ can be deduced straightforwardly for a given coarsening dynamics using scaling arguments \cite{Herring1950,Bray2002}, but it is only possible to predict additional information (such as coarsening rates or morphological evolution) when additional assumptions are made.
In particular, coarsening kinetics and morphological evolution can be determined analytically for low volume fractions of particles that are coarsening via bulk diffusion based on the pioneering work of Lifshitz and Slyozov \cite{Lifshitz1961} and Wagner \cite{Wagner1961} and its extensions (see Ref.\ \cite{Baldan2002} for a review).
However, the approach taken by Lifshitz, Slyozov, and Wagner does not generalize to more complex morphologies, like the bicontinuous structures resulting from spinodal decomposition \cite{Jinnai2000} and dealloying \cite{mccue2016dealloying}, or dendritic structures resulting from solidification \cite{Mendoza2003}, and a comparable theory applicable to these structures has not yet been developed.

As a step toward understanding coarsening of bicontinuous structures, this study considers the question: how does the coarsening mechanism influence the morphological evolution of a bicontinuous structure?
To answer this question, we employ continuum-scale simulations to compare coarsening of bicontinuous structures via two mechanisms, bulk diffusion and surface diffusion.
Both of these mechanisms have been considered independently in previous studies (by Refs.\ \cite{Kwon2007,Kwon2010,Kwon2007thesis} for bulk diffusion and Refs.\ \cite{Geslin2019,li2019topology} for surface diffusion), and we focus our study on the range of volume fractions where bicontinuous structures are known to be stable, namely $\ge 36\%$ of minority phase \cite{Kwon2010,li2019topology}.
We complement the existing information with our own simulation results, which enable comparisons that are not possible based on previous work alone.
We characterize the differences in morphology due to coarsening mechanism and volume fraction, and we focus our discussion on the origins of these differences.
To provide additional context for our observations, we also compare the coarsened structures to two types of model bicontinuous structures: triply periodic area-minimizing surfaces of constant mean curvature \cite{Gozdz1996minimal,Jung2007minimal,pia2017gyroidal} and the level surfaces of random fields generated by superposing waves with a single characteristic wavelength but random phase and orientation \cite{Cahn1965,Teubner1991random,soyarslan20183d}, which we refer to hereafter as `leveled-wave structures' \cite{li2019topology}.

The two coarsening mechanisms we examine, surface and bulk diffusion, are of considerable practical interest.
Coarsening via bulk diffusion occurs in the dendritic solid-liquid systems present during casting \cite{Mendoza2003,Aagesen2010,Aagesen2011}.
Both mechanisms are possible in polymer mixtures undergoing spinodal decomposition \cite{Wolterink2006,Takeno1998}: kinetic power laws in between those of the two coarsening mechanisms ($\ell \propto t^{1/3}$ for bulk diffusion and $\ell \propto t^{1/4}$ for surface diffusion) have been observed experimentally, with behavior closer to surface diffusion farther from the critical temperature.
Coarsening via surface diffusion occurs during dealloying \cite{Qian2007,Wada2011} and annealing \cite{hu2016nanoporous} of nanoporous metals, which are of interest for a variety of functional and structural applications (see \cite{mccue2016dealloying} for a review). 
Nanoporous metals are particularly interesting systems for examining the morphologies that result from coarsening because their strength and compressive stiffness have been shown to be related to their scaled topology (i.e., genus density) \cite{mangipudi2016topology}, one of the metrics we examine.

\section{Background}
In coarsening via both bulk and surface diffusion, mass transport is driven by differences in chemical potential within the structure \cite{Lifshitz1961,Mullins1957}.
We consider coarsening within a two-phase, two-component system with fixed volume in which both species have the same molar volumes in both phases.
Local composition is then fully specified by a single field $\phi(\mathbf x)$, which may be scaled or non-dimensionalized as long as it remains linearly related to the concentrations of each component.
We also assume that the free energies of the bulk phases are zero at equilibrium.
Within this two-phase system, differences in chemical potential arise due to the Gibbs-Thomson effect, which (for isotropic interfacial energy) relates the chemical potential $\mu$ to the local mean curvature $H$ at a point on the interface,
\begin{equation}
\label{eq:gibbs_thomson}
    \mu = \frac{2\gamma H}{\phi \big|_-^+},
\end{equation}
where $\gamma$ is interfacial energy and $\phi \big|_-^+$ represents the difference in $\phi$ between phases.
In coarsening via bulk diffusion, the diffusional flux $\mathbf j$ is proportional to the gradient in chemical potential, $\mathbf j = -M \nabla \mu$, where $M$ is the chemical mobility.
Mass transport is assumed to be at steady state,
\begin{equation}
\label{eq:laplace}
    \nabla^2 \mu = 0,
\end{equation}
and interfacial motion is governed by a mass balance at the interface,
\begin{equation}
\label{eq:bulk_massbal}
    v_n = \frac{ \mathbf n\cdot \mathbf j |_-^+}{\phi \big|_-^+},
\end{equation}
where $\mathbf n$ is the interface normal vector and $v_n$ is the velocity of the interface in the normal direction.
In Eq.\ \ref{eq:bulk_massbal}, any contribution from fluxes within the interface is assumed to be negligible compared to fluxes normal to the interface.
This contrasts with coarsening via surface diffusion, where diffusional fluxes are confined to the interface.
For coarsening via surface diffusion, combining Eq.\ \ref{eq:gibbs_thomson} with equations for the flux and mass balance yields the following expression for the normal velocity of the interface \cite{Mullins1957},
\begin{equation}
    \label{eq:surfdiff}
    v_n = \alpha \nabla^2_S H,
\end{equation}
where $\nabla^2_S$ is the surface Laplacian and $\alpha$ is a kinetic coefficient.
Thus, while both coarsening mechanisms have the same driving force and similar physical mechanisms, evolution via surface diffusion depends only on local interfacial geometry (the surface Laplacian of mean curvature), while evolution via bulk diffusion is affected by non-local geometry via the transport field (the solution to Eq.\ \ref{eq:laplace}).
While the difference in dynamics between coarsening via bulk and surface diffusion can be explained in terms of local versus non-local diffusional interactions, it is not obvious how this difference will affect the morphology of a bicontinuous structure.
Our study, consisting of simulations of coarsening via both mechanisms and characterization of morphological differences between them, is therefore intended as a foundation for future theoretical work into understanding relationships between coarsening mechanism and morphology.

\section{Methods}
\subsection{Phase field model}
To simulate the coarsening of complex structures via the dynamics described in the previous Section, we employ phase field models, which represent interfaces implicitly as a region over which one or more field variables smoothly transition.
The models we use are based on the Cahn-Hilliard equation, which describes the evolution of a single concentration field that represents a two-phase system \cite{CahnHilliard1958, Langer1975CH}.
With constant mobility, the Cahn-Hilliard equation corresponds to phase separation followed by coarsening via bulk diffusion \cite{Pego1989}, and it can be expressed as the generalized diffusion equation \cite{Langer1975CH}
\begin{equation}
\label{eq:CH_diff}
    \frac{\partial \phi}{\partial t} = \nabla \cdot M \nabla \mu,
\end{equation}
where, consistent with their definitions in the previous section, $\phi$ is a scaled concentration and $\mu$ is the chemical potential.
In the Cahn-Hilliard model, the chemical potential $\mu$ is the variational derivative with respect to $\phi$ of the free energy functional \cite{CahnHilliard1958}
\begin{equation}
    \label{eq:CH_freeen}
    F[\phi,\nabla \phi] = \int_{\Omega}{\left[f(\phi)+\frac{\epsilon^2}{2}|\nabla \phi|^2\right] d\mathbf x},
\end{equation}
where $\Omega$ represents the domain, $\epsilon$ is the gradient energy coefficient, a scalar parameter, and the function $f(\phi)$ represents the volumetric free energy density of a bulk (i.e., compositionally homogeneous) phase.
Evaluating this variational derivative to find $\mu$ in Eq.\ \ref{eq:CH_diff} results in
\begin{equation}
\label{eq:CH_eq_constM}
    \frac{\partial \phi}{\partial t} = \nabla \cdot M \nabla \left[ f'(\phi) - \epsilon^2 \nabla^2 \phi \right].
\end{equation}
We additionally specify $f(\phi)$ to be the double-well potential \cite{Cahn1961},
\begin{equation}
\label{eq:doublewell}
    f(\phi) = \frac{W}{4} \phi^2 (1-\phi)^2,
\end{equation}
where the bulk equilibrium concentrations of the phases are $\phi_0^+=1$ and $\phi_0^- = 0$, and the height of the energy barrier between them is set by the parameter $W$.
Our model for coarsening via bulk diffusion thus consists of Eq.\ \ref{eq:CH_eq_constM} with constant mobility $M$ and the bulk free energy $f(\phi)$ given by Eq.\ \ref{eq:doublewell}.

Phase field models for surface diffusion typically include a concentration-dependent mobility that must be nonzero within the interface and zero at the bulk equilibrium concentrations of the two phases, $\phi_0^+$ and $\phi_0^-$  \cite{Lacasta1992,Zhu1999, Cahn1996sharp, Ratz2006}.
However, this approach will still allow some slow bulk diffusion if concentrations in the bulk deviate from $\phi_0^+$ and $\phi_0^-$ \cite{Bray1995, DaiDu2014, Lee2016siam}.
Such deviations in concentration are expected to occur during coarsening due to the Gibbs-Thomson effect, which relates interfacial curvature to chemical potential and thus to concentration \cite{Pego1989}.
R\"atz, Ribalta, and Voigt (RRV) \cite{Ratz2006} proposed a model that reduces these changes in concentration (and the bulk diffusion that they permit) by modifying the chemical potential with a stabilizing function \cite{Ratz2006,Voigt2016comment}.
The RRV model thus converges to coarsening via surface diffusion (Eq.\ \ref{eq:surfdiff}) in the asymptotic limit $\epsilon \to 0$ \cite{Ratz2006,Gugenberger2008}, which corresponds to an asymptotically thin (i.e., sharp) interface.
Salvalaglio et al.\ \cite{salvalaglio2019doubly} have recently proposed an alternative to the RRV model that can be derived from variational principles, unlike the RRV model.
However, their model does not appear to improve convergence to Eq.\ \ref{eq:surfdiff} compared to the RRV model \cite{salvalaglio2019doubly}, and thus we we retain the choice of the RRV model.

The form of the RRV model that we implement consists of Eq.\ \ref{eq:CH_eq_constM} with the following modifications \cite{Ratz2006}: the variationally derived chemical potential is divided by a stabilizing function, $g(\phi) \propto f(\phi)$, and the mobility is made concentration-dependent, with $M(\phi)$ proportional to $f(\phi)$ or $g(\phi)$.
We use $M(\phi)=g(\phi)=\phi^2 (1-\phi)^2$ \cite{Bray1995}.
The resulting governing equation is
\begin{equation}
\label{eq:RRV}
    \frac{\partial \phi}{\partial t} = \nabla \cdot M(\phi) \nabla \frac{1}{g(\phi)} \left[ f'(\phi) - \epsilon^2 \nabla^2 \phi \right].
\end{equation}
By comparing the sharp-interface dynamics of phase field models with and without $g(\phi)$ \cite{Gugenberger2008,Lee2016siam,Voigt2016comment}, it is apparent that $g(\phi)$ eliminates the effect of interfacial curvature on concentration in the bulk due to the Gibbs-Thomson effect, thereby ensuring that $M(\phi)$ is zero in the bulk.
(We consider the effect of $g(\phi)$ on dynamics in greater detail in the Supplemental Material \cite{supplemental}.)
Furthermore, we expect good agreement between Eq.\ \ref{eq:RRV} and its sharp-interface limit during self-similar coarsening because the diffuse interface has reached its steady-state profile and the interfacial width is small relative to the sizes of features within the structure.
Any effects of finite interfacial width on morphology should decrease over the course of our simulations (since the length scales of features are increasing while the width of the interface remains constant), and self-similar evolution implies that diffuse interface effects have become negligible.

Based on the asymptotic analysis of Eq.\ \ref{eq:RRV} in the limit $\epsilon \to 0$, the kinetic coefficient $\alpha$ in the sharp interface model (Eq.\ \ref{eq:surfdiff}) is related to the phase field model parameters by \cite{Ratz2006}
\begin{equation}
    \label{eq:RRV_to_sharp}
    \alpha = \frac{\epsilon}{(\phi_0^+ - \phi_0^-)} \frac{\gamma}{ \int_{\phi_0^-}^{\phi_0^+} g(\phi)d\phi } \int_{\phi_0^-}^{\phi_0^+} { \frac{ M(\phi) }{ \sqrt{2f(\phi)} } d\phi } ,
\end{equation}
where $\gamma$ is the interfacial energy, which is given by
\begin{equation}
\label{eq:interfacial_energy}
    \gamma = \epsilon \int_{\phi_0^-}^{\phi_0^+} \sqrt{2 f(\phi)}d\phi.
\end{equation}
This expression for $\gamma$ is also valid for the bulk diffusion case.
To quantify the resolution of our simulations, we estimate the width of the diffuse interface $L_{int}$ as the ratio of the concentration difference between phases to the maximum slope of the equilibrium 1-D concentration profile $\phi(x)$  \cite{CahnHilliard1958},
\begin{equation}
    L_{int} = \frac{\phi|_-^+}{\max \phi'(x)},
\end{equation}
which for the double-well potential as expressed in Eq.\ \ref{eq:doublewell} reduces to \cite{CahnHilliard1958,Kwon2010}
\begin{equation}
    L_{int} = \frac{\epsilon \phi|_-^+}{\max_{\phi\in [0,1]} \sqrt{2f(\phi)}} = 4\epsilon \sqrt{\frac{2}{W}}.
\end{equation}

\subsection{Simulation parameters and numerical methods}
The RRV model (Eq.\ \ref{eq:RRV}) was used to simulate phase separation and coarsening via surface diffusion.
Parameters of the phase field model were $W=0.4$ and $\epsilon^2 = 0.2$, which correspond to a sharp interface kinetic coefficient $\alpha = 1/6$ and an interfacial width $L_{int}=4$.
Singularity of $1/g(\phi)$ was avoided by the addition of a small positive constant, $\sigma = 10^{-12}$, to the denominator: $1/g(\phi) \approx 1/[g(\phi) +\sigma]$.
Simulations were conducted within a cubic domain, $L_x=L_y=L_z=1024$, with periodic boundary conditions.
The domain was discretized by a uniform Cartesian grid with $\Delta x = 1$, which results in a ratio of interfacial width to grid spacing of $L_{int}/\Delta x = 4$ (compared to $L_{int}/\Delta x = 4$ for Ref.\ \cite{Geslin2019} and $L_{int}/\Delta x = 2$ for Ref.\ \cite{Kwon2010}).
This should result in 4-5 points through the width of an interface that is aligned with a principal grid direction.
Equation \ref{eq:RRV} was solved numerically using the finite difference method.
Explicit (forward Euler) timestepping was used with $\Delta t = 0.04$.
The Laplacian in Eq.\ \ref{eq:RRV} was discretized using the standard seven-point stencil, and the term $\nabla \cdot M(\phi)\nabla \mu$ was approximated by calculating $M(\phi)\nabla \mu$ on half-points and taking a centered difference to find the divergence $\nabla \cdot M(\phi)\nabla \mu$ on grid points.
Simulations of coarsening via bulk diffusion were conducted using Eq.\ \ref{eq:CH_eq_constM} with constant mobility, $M=1$.
The timestep of these simulations was $\Delta t = 0.05$, and other phase field and numerical parameters were the same as the surface-diffusion case.
All simulation parameters are nondimensional.

Simulations were initialized with random noise, with $\phi$ at each grid point sampled from a uniform distribution over the interval $[\left<\phi\right>-0.1,\left<\phi\right>+0.1]$, where $\left<\phi\right>$ denotes the average system composition.
Two average system compositions were examined, $\left<\phi\right>=0.50$ and $\left<\phi\right>=0.36$, which correspond to nominal volume fractions of 50\%-50\% and 36\%-64\%  of the $\phi=1$ and $\phi=0$ phases, respectively.
Since we simulate coarsening via bulk diffusion with the same constant mobility in both phases, the dynamics of both coarsening mechanisms have phase inversion symmetry about $\phi=0.50$.
Simulations with $\left<\phi\right>=0.36$ and $\left<\phi\right>=0.64$ should therefore be statistically equivalent.
Previous simulation studies identify the transition between particulate and bicontinuous structures as being within 30-36\% volume fraction for coarsening via bulk diffusion \cite{Kwon2010} and at 30\% volume fraction for coarsening via surface diffusion \cite{li2019topology}, although Ref.\ \cite{li2019topology} reports the presence of independent particles at up to 40\% volume fraction.
We therefore select $\left<\phi\right>=0.36$ to provide a large contrast to $\left<\phi\right>=0.50$ while still ensuring stable bicontinuous structures with few independent particles.
Effects of coarsening mechanism at lower volume fractions will be considered in future work.

We perform only a single large calculation for each combination of coarsening mechanism and system composition, rather than multiple smaller calculations.
This enables simulations over larger ranges of time and characteristic length scale without the size effect from the periodic boundary conditions.
Simulations were performed on the XSEDE supercomputers Stampede, Stampede2, and Bridges \cite{xsede}.
The $\left<\phi\right>=0.50$ bulk-diffusion dataset has been previously reported on in Ref.\ \cite{andrews2020simulation}.

\subsection{Characterization methods}
After an initial period to allow for phase separation ($t<1.6\times 10^4$ for surface diffusion and $t<1.0\times 10^4$ for bulk diffusion), structures were extracted from the simulation output and characterized.
Statistics of interfacial curvature, topology, autocorrelations of phase and mean curvature, and the structure factor were calculated for each structure, and are reported with the appropriate scaling to evaluate self-similarity of the structures.
The length scale of the simulated structures was primarily quantified by the ratio of domain volume $V$ to total interfacial area $A_T$, $S_V^{-1} = V/A_T$ \cite{Marsh1996}.
We employ an additional length scale $L_p$ that is directly related to $S_V^{-1}$ to scale autocorrelations and structure factors, and we compute a third length scale $\lambda$ from the structure factor to enable a comparison between our results and those of Ref.\ \cite{Geslin2019}.
Details regarding $L_p$ and $\lambda$ are provided in Section \ref{sec:scaling} below.

\subsubsection{Interfacial morphology}
To accurately determine interfacial curvature, concentration data resulting from the simulation was first post-processed using a level-set based method \cite{Park2014}.
The interface of each structure was approximated as a triangular mesh of the $\phi=0.5$ isosurface.
Mean curvature $H$ and Gaussian curvature $K$ were calculated on the simulation grid as described in Ref.\ \cite{Park2014} and interpolated to the centers of mesh triangles.
Surface integrals were discretized using the mesh faces: $\int_S {q dA} = \sum_{i}q_i A_i$, where $A_i$ and $q_i$ correspond to the area of the $i$th mesh face and the value of the integrand $q$ at the center of the $i$th mesh face, respectively.
Likewise, the total interfacial area was calculated as $A_T=\sum_{i} A_i$.
This method was used to calculate statistics of interfacial morphology, including average mean curvature $\left<H\right>$, standard deviation of mean curvature $\sigma_{H}$, and the interfacial shape distribution (ISD).
The ISD \cite{Mendoza2003} is a probability density function for the probability of finding a point on the interface with a specific pair of principal curvatures, $(\kappa_1, \kappa_2)$, which are related to $H$ and $K$ by $H=(\kappa_1 +\kappa_2)/2$ and $K=\kappa_1\kappa_2$.
We express the ISD as
\begin{equation}
\label{eq:ISD}
    P(\kappa_1,\kappa_2) = \frac{1}{A_T} \frac{\partial^2 F_A(\kappa_1,\kappa_2)}{\partial \kappa_1 \partial \kappa_2},
\end{equation}
where $F_A(\kappa_1, \kappa_2)$ is a cumulative area distribution function, defined as the total area within the structure having first and second principal curvatures less than or equal to $\kappa_1$ and $\kappa_2$, respectively.
We discretize the ISD in rotated coordinates $x^\prime = (\kappa_1+\kappa_2)/\sqrt{2}$ and $y^\prime = (\kappa_2-\kappa_1)/\sqrt{2}$ to avoid having bins that are split by the line $\kappa_1=\kappa_2$ (since $\kappa_2 > \kappa_1$ by convention, such bins would consist partially of a `forbidden' region of the ISD).
Bin sizes for the ISDs are $\Delta x^\prime/S_V = \Delta y^\prime/S_V = 0.125$.
The sign convention for mean curvature is such that convex bodies (e.g., spheres) of the $\phi=1$ phase have positive mean curvature.

\subsubsection{Topology}
Topology of the structures was primarily evaluated in terms of the genus density $g_V$ of the $\phi = 1$ phase, the ratio of genus $g$ to domain volume $V$, $g_V = g/V$.
In general, the topology of a phase in 3D can be quantified by the Betti numbers, $\beta_0$, the number of independent particles/regions of phase, $\beta_1$ (equivalent to the genus, $g$), the total number of `handles' if the particles are deformed into spheres with handles, and $\beta_2$, the number of internal voids or cavities enclosed in the phase \cite{Odgaard1993}.
The number of independent particles $\beta_0$ was calculated using the LABEL\_REGION function in IDL and an algorithm to take into account connections through the periodic boundaries.
The genus was evaluated by calculating the Euler characteristic $\chi$ and applying the Euler-Poincar\'e formula,
\begin{equation}
\label{eq:euler}
    \chi = \beta_0-\beta_1+\beta_2,
\end{equation}
where it is assumed that $\beta_2=0$.
(We do not check this assumption explicitly, although $\beta_2$ should be statistically equivalent to $\beta_0$ in the $\left<\phi\right>=0.50$ cases due to phase-inversion symmetry.)
The Euler characteristic of the interface mesh, $\chi_m$ (twice the Euler characteristic of each phase, $\chi_m=2\chi$) was calculated with the Euler-Poincar\'e formula, $\chi_m = N_V - N_E + N_F$, where $N_V$ is the number of vertices, $N_E$ the number of edges, and $N_F$ the number of faces \cite{MWEuler}.
The formula for the genus is then $g = \beta_0 - \chi_m/2$.

We calculate $g$ as an extensive property of a structure, such that a larger structure that contains $n$ unit cells will have a genus of $ng$.
For $g$ to be an extensive property, we must avoid over-counting topological features that are shared by multiple unit cells.
Therefore, we exclude one half of the contributions from edges and vertices on domain faces to $\chi_m$ in the Euler-Poincar\'e formula  \cite{Odgaard1993}.
When calculating $\beta_0$, we take advantage of the periodicity of the structures to determine whether regions of the $\phi=1$ phase are connected to the main bicontinuous region of the $\phi=1$ phase through domain boundaries.
We also exclude the bicontinuous region itself from $\beta_0$, since it is shared between all unit cells.
Thus, a periodic bicontinuous structure has $\beta_0=0$ when evaluated by our method.
Genus density is related to Gaussian curvature $K$ via the Gauss-Bonnet theorem, which is expressed as
\begin{equation}
    \label{eq:GaussBonnet}
    g_V = -\frac{1}{4\pi V} \int_S {K dA}
\end{equation}
for a bicontinuous structure.

\subsubsection{Autocorrelations and structure factor}
Two-point statistics methods can be used to capture information about the spatial distribution of features in a microstructure, unlike other descriptors that treat each data point as an independent sample.
Recent work \cite{Sun2017,Sun2018, Sun_2019} extends on methods that only provide information about the correlation between quantities in the bulk of the microstructure to also include interfacial quantities, such as the mean curvature.
Of particular interest is the two-point Pearson correlation function, which quantifies the degree of correlation between two spatial fields and allows identification of both positive and negative correlations.
The discretized two-point Pearson correlation is defined as
\begin{equation}
\label{eq:pearson}
    h[u,u';\mathbf{q}] = \frac{1}{|\mathbb{S}[\mathbf{q}]|} \frac{1}{|\sigma[\mathbf{q}]\sigma'[\mathbf{q}]|} \sum_{\mathbf{p} \in \overline{\mathbb{S}}} \left(u[\mathbf{p}]-\bar{u} \right)c[\mathbf{p}] \cdot \left(u'[\mathbf{p}+\mathbf{q}]-\bar{u}'\right) c[\mathbf{p}+\mathbf{q}]
\end{equation}
where $u$ and $u'$ are local attributes of interest in the microstructure, such as the interfacial curvature; for autocorrelation, $u=u'$.
The symbol $\mathbf{q}$ denotes the vector defining the separation of the two points being correlated, $\mathbf{p}$ is the index of the spatial bins, $\mathbb{S}$ is the entire set of spatial bins, and $|\mathbb{S}[\mathbf{q}]|$ is the total number of spatial bins in $\mathbb{S}[\mathbf{q}]$.
The total number of spatial bins in $\mathbb{S}[\mathbf{q}]$ is given by
\begin{equation}
|\mathbb{S}[\mathbf{q}]| = \sum_{\mathbf{p} \in \overline{\mathbb{S}}} c[\mathbf{p}]c[\mathbf{p}+\mathbf{q}]
\end{equation}
where $c$ is a mask function, which tracks the voxels that should be included in the statistics calculation.
In the case of a cubic periodic data set, this mask function will include the entire data set when $u$ is a quantity in the bulk and will only include voxels on the interface when $u$ is an interfacial quantity.
The standard deviations of $u$ and $u'$ are given by $\sigma$ and $\sigma'$ and their means are given by $\bar{u}$ and $\bar{u}'$.
Further mathematical details are provided in Ref.\ \cite{Sun2017}. The two-point Pearson correlations range from -1 to 1, where $h[u,u';\mathbf{q}]=1$ is perfect correlation, $h[u,u';\mathbf{q}]=-1$ is perfect anti-correlation, and $h[u,u';\mathbf{q}]=0$ is no correlation.
The evolution of the interface is controlled primarily by $H$, the interfacial mean curvature, thus two-point Pearson autocorrelations will be calculated of $H$.
We shall also determine the spatial autocorrelation of the segmented phases (i.e., the phase autocorrelations), where voxels are binarized to values of zero or one if $\phi \le 0.5$ or $\phi > 0.5$, respectively.

To enable a comparison between our work and the results of Geslin et al.\ \cite{Geslin2019} for surface diffusion, we also calculate the structure factor of our coarsened structures.
The structure factor is useful since it can be directly obtained from diffraction data ~\cite{Bray2002,debye1957scattering,Teubner1990}.
The structure factor is a function of wave vector $\mathbf k$, which we define such that the wavenumber $|\mathbf k|=k$ is the inverse of the wavelength (i.e., without the factor of $2\pi$ that is sometimes included \cite{Geslin2019}).
The structure factor $S(\textbf{k)}$ is then defined as \cite{Bray2002,Kwon2007thesis}
\begin{equation}
\label{eq:structurefact}
    S(\textbf{k})=\frac{1}{V}\hat{\phi}(\textbf{k})\hat{\phi}^{*}(\textbf{k})
\end{equation}
where $V$ is the total volume of the domain, $\hat{\phi}^*$ denotes the complex conjugate of $\hat{\phi}$, and $\hat{\phi}$ is the Fourier transform of the concentration $\phi$,
\begin{equation}
\label{eq:FT}
    \hat{\phi}(\textbf{k})=\int_{\Omega} \phi(\textbf{k})e^{-2\pi i \textbf{k} \cdot \mathbf{x}}d\mathbf x
\end{equation}
where $\mathbf{x}$ is the position vector.

For computational efficiency without loss of accuracy, the data is downsampled from $1024^3$ to $512^3$ voxels for calculation of the structure factor and autocorrelations.
As the simulated dynamics are isotropic, radial averages of the autocorrelations and structure factors are calculated as functions of the radial distance in real space or $k$-space, respectively.
The radial average of $h[u,u';\mathbf{q}]$ or $S(\textbf{k})$ is thus the average over a sphere with radius $r$ or $k$, respectively, centered at the origin.
The radial averages were calculated in equidistant bins with sizes $\Delta r = 2$ (i.e., one voxel-width in the downsampled structure) and $\Delta k = 2/L_x=1/512$.

\subsubsection{Scaling\label{sec:scaling}}
We employ $S_V^{-1}$ to scale curvatures and genus density.
This is a natural choice of length scale for coarsening since it is inversely proportional to the total interfacial energy $\gamma A_T$ and thus directly connected to the thermodynamics of the system.
With $S_V^{-1}$ as the characteristic length scale, scaled mean curvature is $H/S_V$, scaled Gaussian curvature is $K/S_V^2$, scaled genus density is $g_V S_V^{-3}$, and the ISDs are reported in terms of scaled principal curvatures $\kappa_1/S_V$ and $\kappa_2/S_V$.
The Gauss-Bonnet theorem (Eq.\ \ref{eq:GaussBonnet}) expressed in terms of scaled quantities is
\begin{equation}
    \label{eq:GaussBonnetsc}
    -\frac{1}{4\pi}\frac{\int_S {K S_V^{-2} dA}}{A_T} = g_V S_V^{-3}.
\end{equation}
That is, the scaled genus density $g_V S_V^{-3}$ is proportional to the average of the scaled Gaussian curvature $K/S_V^2$ over the interface.

To scale radial distance in the structure factor and phase and $H$ autocorrelations, we employ a length scale based on Porod's Law \cite{porod1951,debye1957scattering}, which states that the first derivative of $h(r)$, the radially averaged phase autocorrelation, at $r=0$ is related to $S_V^{-1}$ and the volume fraction, $V_f$, by
\begin{equation}
    \label{eq:porodlaw}
    h'(0) = - \frac{1}{4  V_f(1-V_f) S_V^{-1}}
\end{equation}
Based on Eq.\ \ref{eq:porodlaw}, scaling $r$ by the length scale $L_p = 4  V_f(1-V_f) S_V^{-1}$ ensures that all of the phase autocorrelations have the same slope at $r=0$.
They will have the same value at $r=0$ as well, since $h(0)=1$ for any Pearson autocorrelation.
Thus, scaling $r$ by $L_p$ ensures that the phase autocorrelations of all cases have the same behavior near $r=0$.
A closely related prediction also exists for the large-$k$ behavior of the structure factor of a segmented structure \cite{porod1951,Teubner1990},
\begin{equation}
\label{eq:porodstruc}
    S(k) \approx \frac{1}{8\pi^3 V_f(1-V_f) S_V^{-1} k^4} = \frac{1}{2 \pi^3 L_p k^4},
\end{equation}
which suggests that scaling the wavenumber $k$ by $1/L_p$ and $S(k)$ by $L_p^3$ will ensure that structure factors have the same behavior in the limit $k\to \infty$ \cite{Bray2002}.
By accounting for the known effects of volume fraction in Eqs.\ \ref{eq:porodlaw} and \ref{eq:porodstruc}, these scalings by $L_p$ allow us to test for more subtle differences between autocorrelations and structure factors for different cases.

We compute an additional length scale $\lambda$ associated with the structure factor to enable comparisons between our results and those of Geslin et al.\ \cite{Geslin2019} for coarsening via surface diffusion.
Instead of $S_V^{-1}$ or $L_p$, Geslin et al.\ chose to scale microstructural quantities by the wavelength $\lambda=1/k_0$, where $k_0$ is the first moment of the structure factor,
\begin{equation}
    k_0 = \frac{\int_0^\infty k S(k) dk}{\int_0^\infty S(k)}.
    \label{eq:k0}
\end{equation}

\subsection{Model structures\label{sec:models}}
We compare our coarsened structures to two types of model structure, constant-mean-curvature surfaces and leveled-wave structures.
Jung et al.\ \cite{Jung2007minimal} generated constant-mean-curvature surfaces with the same symmetries as the Schwarz P, Schwarz D and Schoen G minimal surfaces over a range of volume fractions using an area-minimization algorithm.
For each volume fraction and type of surface, Jung et al.\ report the area $A_T$ and constant mean curvature $H$ for a unit cell of the surface embedded in a cube with unit volume.
From these values, we calculate $S_V$, $\left<H/S_V\right>$, and $g_V S_V^{-3}$ (where the values of genus per unit cell are known to be 3, 9, and 5 for the Schwarz P, Schwarz D and Schoen G families of surfaces, respectively \cite{Gozdz1996minimal}).
As an example, Jung et al.\ report $S_V=3.04$ and $H=0.73$ for the Schoen G surface at $V_f=0.4$, which result in $g_V S_V^{-3}=5/3.04^3=0.178$ and $\left<H/S_V\right>=0.73/3.04=0.24$.
Since Jung et al.\ \cite{Jung2007minimal} tabulate values at increments of $0.05$ in $V_f$, we calculate $g_V S_V^{-3}$ and $\left<H/S_V\right>$ directly from their data at $V_f=0.50$ and obtain values at $V_f=0.36$ by cubic interpolation.
For brevity, we compare only the Schoen G family of surfaces to our coarsened structures in the main text, and the data justifying this choice (a comparison with the Schwartz D and P surfaces) is given in the Supplemental Material \cite{supplemental}.

The leveled-wave model structure that we consider consists of thresholded Gaussian random fields generated by superposing waves with random phase and orientation and a single fixed wavelength \cite{Cahn1965,soyarslan20183d}.
For this structure, Soyarslan et al.\ \cite{soyarslan20183d} provide analytical expressions for $\left<H \right>$ and $g_V$ as functions of the input wavenumber and a normalized thresholding parameter $\xi$, which is related to volume fraction by $\xi = \sqrt{2} \mathrm{erf}^{-1}(2 V_f -1)$ ($\mathrm{erf}^{-1}$ denotes the inverse error function).
Soyarslan et al.\ also provide an analytical relationship between the input wavenumber and $S_V$ that allows us to express the scaled quantities $g_VS_V^{-3}$ and $\left<H/S_V \right>$ in terms of $\xi$ alone,
\begin{equation}
\label{eq:hsv_xi}
    \left<H/S_V \right> = - \frac{1}{2}\left( \frac{\pi}{2} \right)^{3/2}  \xi e^{\xi^2/2},
\end{equation}
and
\begin{equation}
\label{eq:gvsv_xi}
    g_V S_V^{-3} = \frac{\pi}{32}  \left(1-\xi^2\right)e^{\xi^2}.
\end{equation}
Based on a result from Ref.\ \cite{Teubner1991random}, we also derive an expression for the standard deviation of scaled mean curvature $\sigma_{H/S_V}$ as a function of $\xi$ for the leveled-wave model structure,
\begin{equation}
\label{eq:sigh_xi}
    \sigma_{H/S_V} = \frac{\pi}{2\sqrt{2}} \left[  \left(1- \frac{\pi}{4} \right) \xi^2  +  \frac{1}{5}  \right]^{1/2}  e^{\xi^2/2}.
\end{equation}
Details of the derivations of Eqs.\ \ref{eq:hsv_xi}, \ref{eq:gvsv_xi}, and \ref{eq:sigh_xi} are given in the Supplemental Material \cite{supplemental}.

\section{Results and Discussion}
In this section, we show how the scaled morphology and topology of the different structures converge to self-similar states, and how the kinetics of the surface-diffusion and bulk-diffusion cases converge respectively to the $t^{1/4}$ and $t^{1/3}$ power laws predicted for self-similar coarsening.
We then analyze the self-similarly evolving structures in greater detail and discuss possible explanations for the differences we observe between surface and bulk diffusion.

\subsection{Convergence to self-similar coarsening}
\subsubsection{\label{sec:evolution}Evolution of scaled morphology and topology}
\begin{figure}
	\centering
	\includegraphics{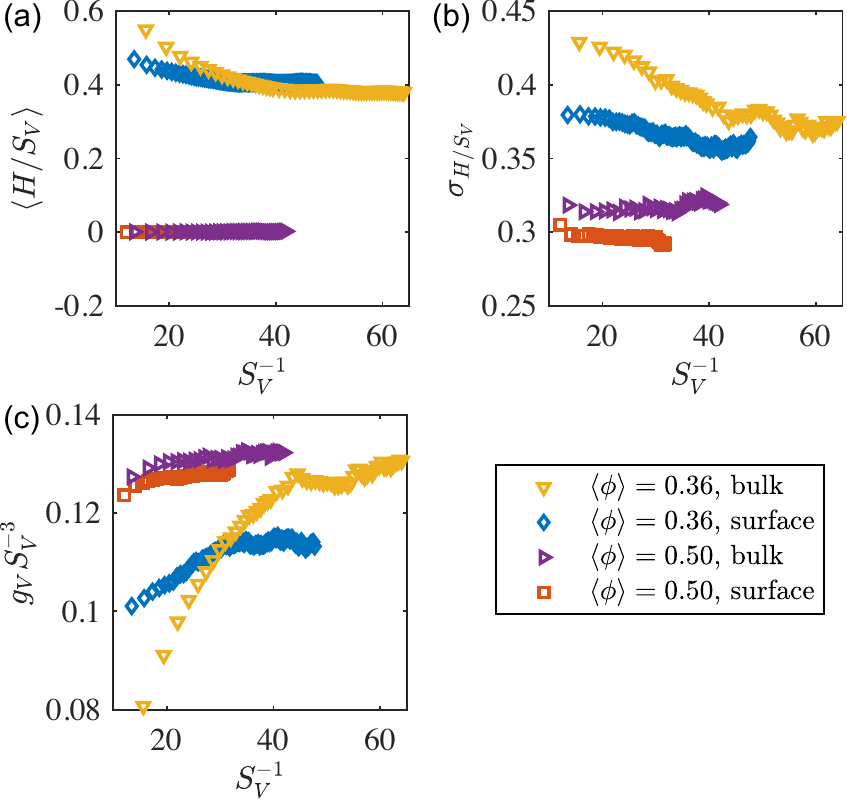}
	\caption{
	Evolution of the bulk- and surface-diffusion structures characterized by (a) average scaled mean curvature $\left< H/S_V\right>$, (b) standard deviation of scaled mean curvature $\sigma_{H/S_V}$, and (c) scaled genus density $g_V S_V^{-3}$.  These quantities are plotted against the characteristic length $S_V^{-1}$ for all four cases: bulk diffusion $\left< \phi \right> =0.36$ (yellow downward-pointing triangles), surface diffusion $\left< \phi \right> =0.36$ (blue diamonds), bulk diffusion $\left< \phi \right> =0.50$ (purple rightward-pointing triangles), and surface diffusion $\left< \phi \right> =0.50$ (red squares).
	}
	\label{fig:evolution}
\end{figure}

Figure \ref{fig:evolution} shows the average scaled mean curvature $\left< H/S_V\right>$, standard deviation of scaled mean curvature $\sigma_{H/S_V}$, and scaled genus density $g_V S_V^{-3}$ plotted against the characteristic length, which enables a direct comparison between the morphological evolution resulting from surface and bulk diffusion despite their different kinetics.
In the structures with $\left< \phi \right> =0.50$, scaled genus density converges quickly, and the average and standard deviation of mean curvature (Fig.\ \ref{fig:evolution}a and \ref{fig:evolution}b) undergo little if any evolution.
This indicates that the morphologies resulting from phase separation are close to the self-similar coarsening morphologies.
More substantial evolution is observed at $\left< \phi \right> =0.36$; for both coarsening mechanisms, the average and standard deviation of scaled mean curvature decrease while scaled genus density increases.
In all four cases (bulk and surface diffusion with $\left< \phi \right> =0.50$ and $\left< \phi \right> =0.36$), there is a late time/large $S_V^{-1}$ regime during which all of the quantities plotted in Fig.\ \ref{fig:evolution} appear to be stable.

To define a converged regime for each structure for further analysis, we applied an ad hoc convergence criterion based on the evolution of $g_V S_V^{-3}$ with respect to $S_V^{-1}$ (i.e., the series plotted in Fig.\ \ref{fig:evolution}c).
We focus on $g_V S_V^{-3}$ since the cases with $\left< \phi \right> =0.50$ have transient behavior in $g_V S_V^{-3}$ that is less apparent in $\left< H/S_V\right>$ or $\sigma_{H/S_V}$.
Our criterion resulted in converged regimes starting from $S_V^{-1}=19.1$ and $S_V^{-1}=32.4$ for the surface-diffusion $\left<\phi\right>=0.50$ and $\left<\phi\right>=0.36$ structures, respectively, and $S_V^{-1}=21.5$ and $S_V^{-1}=46.7$ for the bulk-diffusion $\left<\phi\right>=0.50$ and $\left<\phi\right>=0.36$ structures, respectively. 

\begin{figure}
	\centering
	\includegraphics{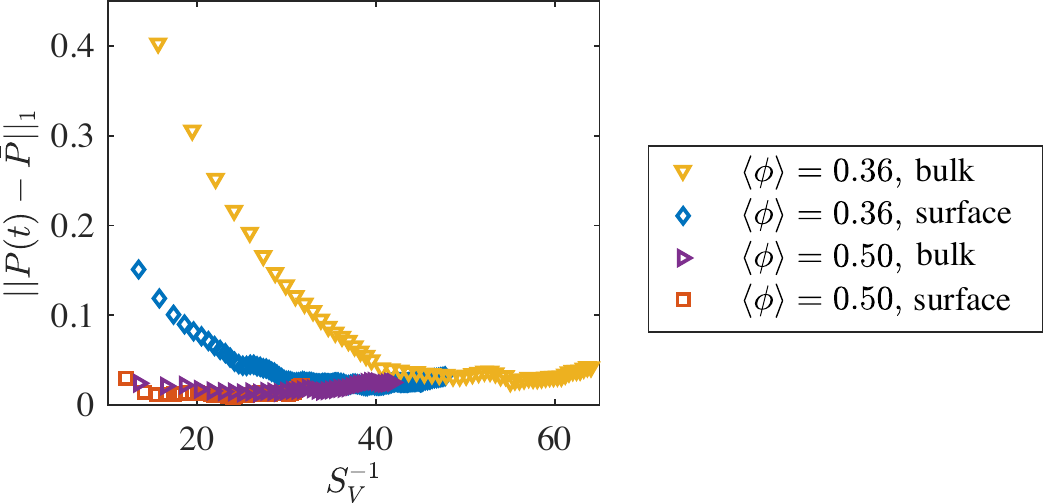}
	\caption{
	Plot of $||P(t)-\bar P||_1$, a measure of the convergence of the ISD, against $S_V^{-1}$. $||P(t)-\bar P||_1$ is the integrated absolute difference between the converged, time-averaged ISD $\bar P$ and the time-dependent ISD $P(t)$.  All four conditions are shown: bulk diffusion $\left< \phi \right> =0.36$ (yellow downward-pointing triangles), surface diffusion $\left< \phi \right> =0.36$ (blue diamonds), bulk diffusion $\left< \phi \right> =0.50$ (purple rightward-pointing triangles), and surface diffusion $\left< \phi \right> =0.50$ (red squares).
	}
	\label{fig:ISDconv}
\end{figure}

To assess whether evolution is in fact self-similar in these regimes, we examine the difference between an average ISD over the converged regime, denoted by $\bar P$, and the ISD at each output step, $P(t)$.
To construct $\bar P$ to be representative of the converged state with reduced statistical variation, we calculate it for each simulation using ISDs from all of the simulation's outputs within the temporal range corresponding to its converged regime.
Weights for this average are determined by the changes in $S_V^{-1}$ between outputs, which ensures that outputs are weighted by the amount of evolution (and thus variation) that they represent.
Figure \ref{fig:ISDconv} plots the integrated absolute difference between $P(t)$ and $\bar P$ (i.e., the $L^1$ normed difference $||\bar P - P(t)||_1$) against the characteristic length.
The integral of an ISD (equal to its $L^1$ norm $||P||_1$) is normalized to unity as it is a probability density function, so the difference $||\bar P - P(t)||_1$ is the relative difference between the time-averaged and evolving ISDs.
For example, $P(t)$ at the earliest simulation output differs by $40\%$ from $\bar P$ for the bulk-diffusion case with $\left<\phi\right>=0.36$.
For the $\left<\phi\right>=0.50$ cases, the difference $||\bar P - P(t)||_1$ is always small, which is consistent with the minimal evolution of $\left< H/S_V\right>$ and $\sigma_{H/S_V}$ in Fig.\ \ref{fig:evolution}.
For the $\left<\phi\right>=0.36$ cases, $||\bar P - P(t)||_1$ decreases substantially during coarsening to low, stable values, indicating convergence of the ISD.
A slight gradual increase in $||\bar P - P(t)||_1$ is observed in each case in Fig.\ \ref{fig:ISDconv} after convergence, a phenomenon that we consider in greater detail in the Supplemental Material \cite{supplemental}.
The Supplemental Material also contains information regarding the convergence of the autocorrelations of phase and mean curvature.

\subsubsection{Kinetics}
During self-similar coarsening, the time evolution of the characteristic length $S_V^{-1}$ must follow the appropriate power law for each coarsening mechanism.
These coarsening power laws can be expressed as
\begin{equation}
    \label{eq:powerlawsurf}
    S_V^{-4}(t) = k_s (t-t_{0})
\end{equation}
for surface diffusion and
\begin{equation}
    \label{eq:powerlawbulk}
    S_V^{-3}(t) = k_b (t-t_0)
\end{equation}
for bulk diffusion, where $k_s$ and $k_b$ are respectively the coarsening rate constants for surface and bulk diffusion, $t$ is the simulation time, and the time offset $t_0$ represents a constant of integration that accounts for finite $S_V^{-1}$ and/or $t$ at the onset of power-law coarsening.
The time offset $t_0$ and the rate constants $k_s$ and $k_b$ are not known analytically for bicontinuous structures and must be determined empirically (e.g., by fitting).

\begin{figure}
	\centering
	\includegraphics{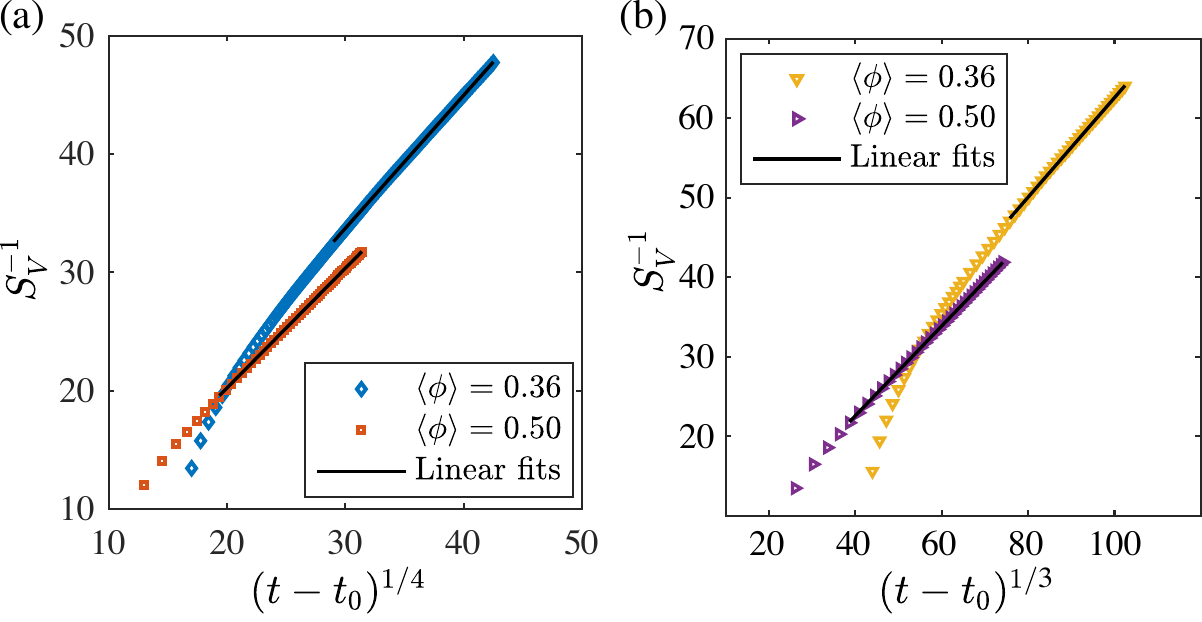}
	\caption{
		Kinetics of coarsening via (a) surface diffusion at $\left< \phi \right> =0.36$ (blue diamonds) and $\left< \phi \right> =0.50$ (red squares) and (b) bulk diffusion at $\left< \phi \right> =0.36$ (yellow downward-pointing triangles) and $\left< \phi \right> =0.50$ (purple rightward-pointing triangles).  Characteristic length $S_V^{-1}$ is plotted against $(t-t_0)^{1/4}$ in (a) and $(t-t_0)^{1/3}$ in (b), where $t_0$ is a fitted constant determined independently for each condition.  Fits of data within the converged regime for each case to the appropriate kinetic power law, Eq.\ \ref{eq:powerlawsurf} in (a) and Eq.\ \ref{eq:powerlawbulk} in (b), are shown as solid black lines.
	}
	\label{fig:kinetics_both}
\end{figure}

Figures \ref{fig:kinetics_both}a and b depict the evolution of the characteristic length $S_V^{-1}$  for the surface- and bulk-diffusion cases, respectively.
Linear fits to the appropriate kinetic power law (Eq.\ \ref{eq:powerlawsurf} for surface diffusion and Eq.\ \ref{eq:powerlawbulk} for bulk diffusion) that are calculated from data within the converged regime for each case are shown as solid black lines.
To show how coarsening kinetics converge to the power laws, Fig.\ \ref{fig:kinetics_both}a plots $S_V^{-1}$ against $(t-t_0)^{1/4}$ for the surface-diffusion cases, and Fig.\ \ref{fig:kinetics_both}b plots $S_V^{-1}$ against $(t-t_0)^{1/3}$ for the bulk-diffusion cases.
The equations of fit for the surface-diffusion cases in Fig.\ \ref{fig:kinetics_both}a are $S_V^{-4} = 1.59t + 1.08\times 10^5$ for $\left< \phi \right>=0.36$ and $S_V^{-4} = 1.05t + 1.30\times 10^4$ for $\left< \phi \right>=0.50$, with coefficients of correlation $R^2 = 0.99986$ and $R^2 = 0.99998$, respectively.
The equations of fit for the bulk-diffusion cases in Fig.\ \ref{fig:kinetics_both}b are $S_V^{-3} = 0.245t + 1.84\times 10^4$ ($R^2 = 0.99960$) for $\left< \phi \right>=0.36$ and $S_V^{-3} = 0.180t + 1.40\times 10^3$ ($R^2 = 0.99996$) for $\left< \phi \right>=0.50$.
In all four cases, the coefficients of correlation near unity indicate excellent agreement with the appropriate power law for each coarsening mechanism (Eq.\ \ref{eq:powerlawsurf} for surface diffusion or Eq.\ \ref{eq:powerlawbulk} for bulk diffusion) within the converged regime, which is a necessary condition for self-similar evolution.

Prior to the converged regime, transient kinetics are observed: the coarsening rate constant decreases over time to its self-similar value in all four cases.
Transient coarsening kinetics can be interpreted as evolution of the coarsening rate constants, where $k_s(t)=dS_V^{-4}/dt$ and $k_b(t)=dS_V^{-3}/dt$.
The observed decreases in coarsening rate constants are larger for the $\left< \phi \right> =0.36$ cases, which correlates to their more substantial evolution prior to convergence in Fig.\ \ref{fig:evolution}.
For coarsening of bicontinuous structures via bulk diffusion, the coarsening rate constant $k_b$ has been found to be linearly related to $\sigma^2_{H/S_V}$, the variance of scaled mean curvature \cite{andrews2020simulation}.
Thus, the decrease in coarsening rate in the $\left< \phi \right> =0.36$ case in Fig.\ \ref{fig:kinetics_both}b should correspond to the decrease in $\sigma_{H/S_V}$ observed in Fig.\ \ref{fig:evolution}b.
This relationship between $\sigma^2_{H/S_V}$ and $k_b$ also explains the higher coarsening rate at $\left< \phi \right> =0.36$ compared to $\left< \phi \right> =0.50$.
In the surface-diffusion cases, we see a qualitatively similar relationship between $\sigma_{H/S_V}$ and $k_s$, with decreasing $\sigma_{H/S_V}$ and $k_s$ during transient coarsening at $\left< \phi \right> =0.36$, and higher $k_s$ and $\sigma_{H/S_V}$ at $\left< \phi \right> =0.36$ compared to  $\left< \phi \right> =0.50$. 
This trend has been noted previously Geslin et al.\ \cite{Geslin2019}, who explain it by noting that variation in curvature leads to coarsening.

\subsection{Self-similar morphologies}
We now compare self-similar morphologies from the surface- and bulk-diffusion cases.
We expect that these morphologies represent those that are reached after a sufficient amount of evolution from a range of initial morphologies.
Therefore, the comparison should be free of the effect of the initial morphologies generated by the different phase separation dynamics, isolating the effect of coarsening mechanism for a given system composition/volume fraction.

\subsubsection{Structures}
\begin{figure*}
	\centering
	\includegraphics{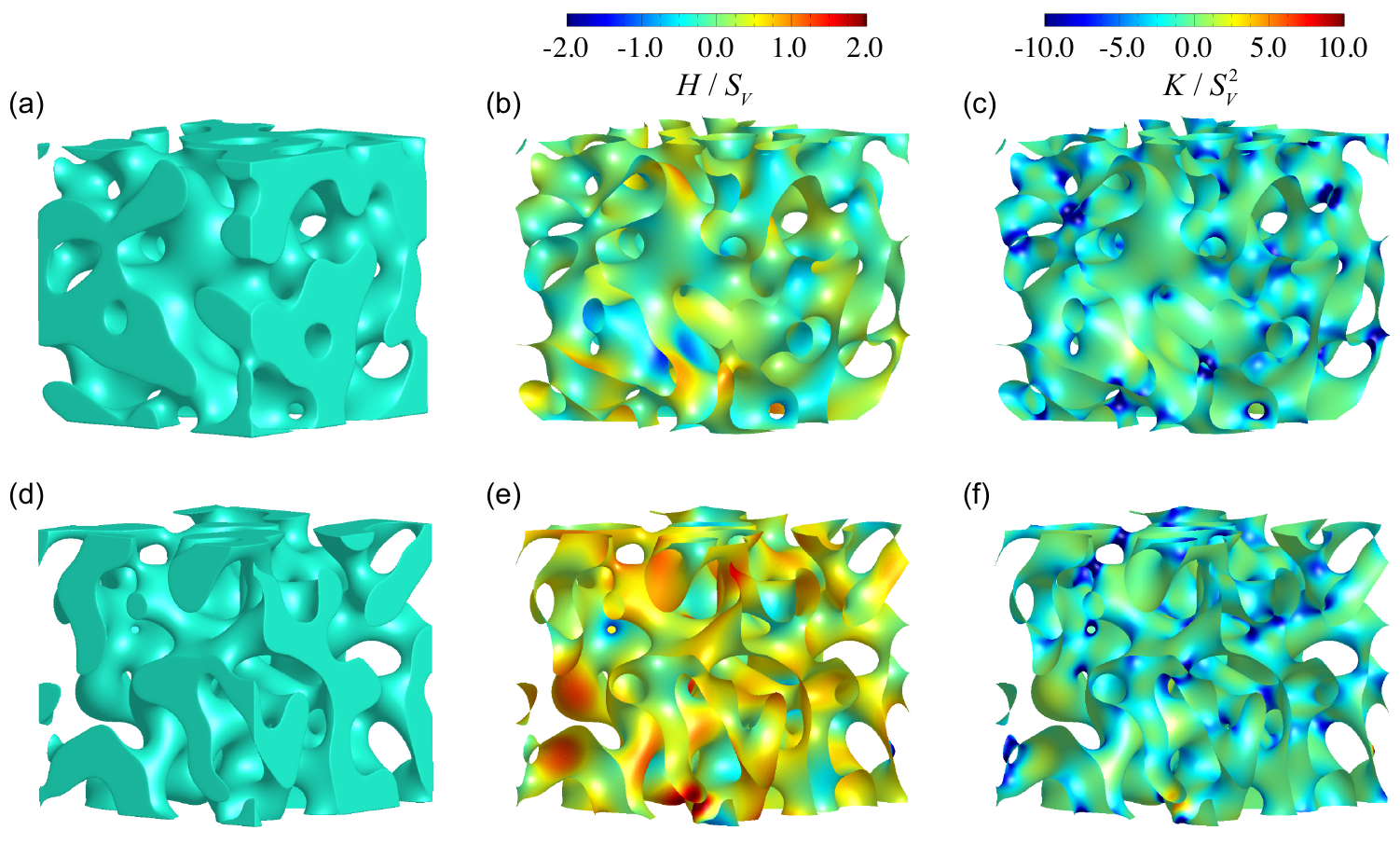}
	\caption{
	Small portions of the simulation volume (of size $(8S_V^{-1})^3$) of structures with $\left<\phi\right>=0.50$ (top row) and $\left<\phi\right>=0.36$ (bottom row), both evolved via surface diffusion.  The $\phi = 0.5$ isosurface is shown in (a,d) with the $\phi=1$ phase capped at the domain boundary, in (b,e) colored by scaled mean curvature, and in (c,f) colored by scaled Gaussian curvature.
	}
	\label{fig:structures}
\end{figure*}

To illustrate qualitative features of the self-similarly evolving structures, Fig.\ \ref{fig:structures} shows samples of the surface-diffusion structures from the latest available times with the same scaled volume, $(8S_V^{-1})^3$.
The $\left<\phi\right>=0.50$ structure at $t=9.6\times 10^5$ is shown in Figs.\ \ref{fig:structures}a-c, and the $\left<\phi\right>=0.36$ structure at $t=3.2\times 10^6$ is shown in Figs.\ \ref{fig:structures}d-f.
The interfaces (i.e., the $\phi=0.50$ isosurfaces) are depicted with the $\phi=1$ phase capped at the sample boundary in Figs.\ \ref{fig:structures}a and \ref{fig:structures}d, colored by scaled mean curvature  $H/S_V$ in Figs.\ \ref{fig:structures}b and \ref{fig:structures}e, and colored by scaled Gaussian curvature $K/S_V^2$ in Figs.\ \ref{fig:structures}c and \ref{fig:structures}f.

Overall, the $\left<\phi\right>=0.36$ structure has more area with positive $H/S_V$ than the $\left<\phi\right>=0.50$ structure (Figs.\ \ref{fig:structures}b and \ref{fig:structures}e, respectively).
This is a direct consequence of the reduced volume fraction of the $\phi=1$ phase in the $\left<\phi\right>=0.36$ case, as it leads to more compact structures of the $\phi=1$ phase and, by our sign convention, convex regions of the $\phi=1$ phase have positive mean curvature.
In Figs.\ \ref{fig:structures}c and \ref{fig:structures}f, most features having large $|K|$ are necks, which have highly negative Gaussian curvature and mean curvature of either sign.
Interfaces with negative Gaussian curvature are hyperbolic interfaces, and their principal curvatures $\kappa_1$ and $\kappa_2$ have opposite signs.
During coarsening, necks shrink and pinch off, leaving behind cones, which rapidly flatten into caps with highly positive or negative mean curvature and positive Gaussian curvature \cite{Bouttes2016,andrews2020simulation,Aagesen2010,Aagesen2011}.
Some caps with positive mean curvature can be seen in the $\left<\phi\right>=0.36$ structure (Figs.\ \ref{fig:structures}e and \ref{fig:structures}f).
In general, interfaces with positive Gaussian curvature are elliptic interfaces, and their principal curvatures have the same sign.
While more elliptic interfaces are visible in the $\left<\phi\right>=0.36$ structure than in the $\left<\phi\right>=0.50$ structure, they are less common than hyperbolic interfaces in both structures.

Both of the structures in Fig.\ \ref{fig:structures} appear to be bicontinuous.
All 40 simulation outputs for each of the $\left< \phi \right>=0.50$ structures were confirmed to be bicontinuous regardless of the coarsening mechanism; while we cannot rule out a presence of an independent domain between output steps, no independent particles were detected in any of the simulated structures for which data was produced.
In simulations at $\left< \phi \right>=0.36$, independent particles were observed at early times for both coarsening mechanisms.
In the bulk-diffusion structure at $\left< \phi \right>=0.36$, all particles disappeared prior to the converged regime of evolution, while in the surface-diffusion structure, two independent particles persisted to the end of the simulation.
These particles of the $\phi=1$ phase comprise a negligible fraction of the total volume, $V_f < 10^{-4}$, and contribute negligibly to the overall topology of the structure, which has $g\ge 1116$ for all simulation outputs, and.
These isolated particles cannot evolve by surface diffusion after reaching their equilibrium shape unless they rejoin the main region of phase.
Therefore, the evolution of the $\left< \phi \right>=0.36$ surface-diffusion case corresponds to that of a bicontinuous structure consisting of its continuous regions of the $\phi=0$ and $\phi=1$ phases.

\subsubsection{Structure factor and two-point Pearson autocorrelations}
\begin{figure}
	\centering
	\includegraphics{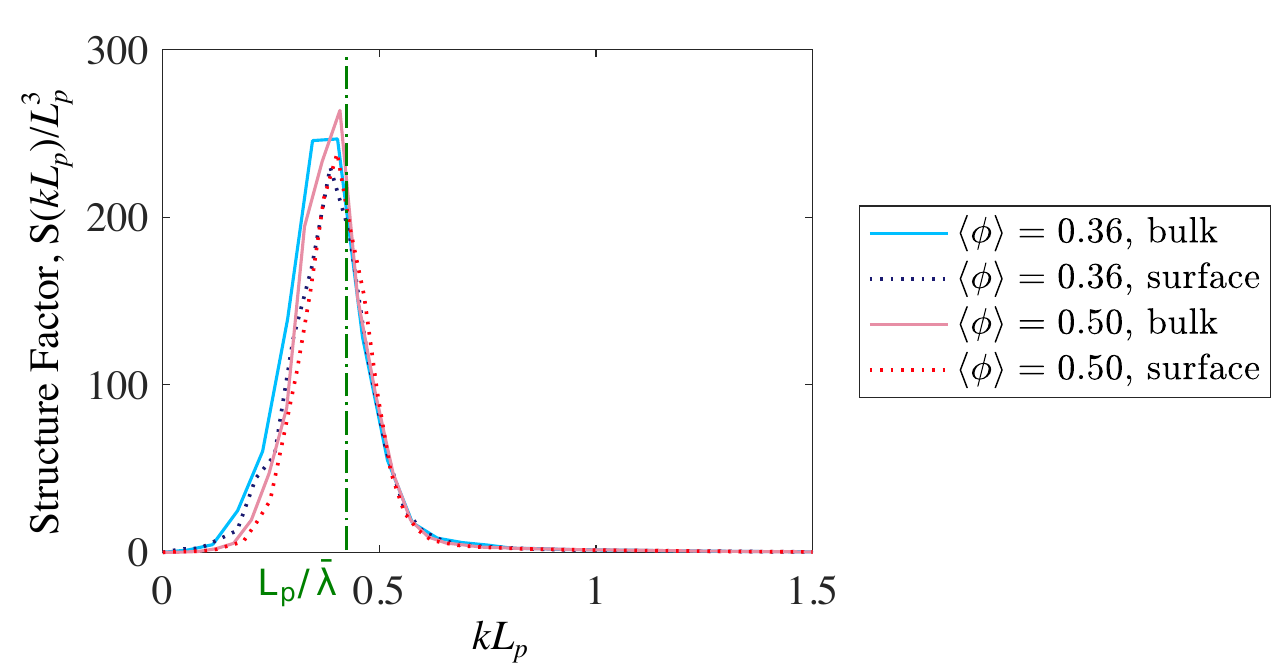}
	\caption{
Structure factors for the $\left<\phi\right>=0.36$ structure with evolution governed by bulk diffusion (cyan line) and surface diffusion (black dashed line) and for the $\left<\phi\right>=0.50$ structures with evolution governed by bulk diffusion (light magenta line) and surface diffusion (red dashed line). The green dash-dot line represents $L_p/\bar \lambda$, the value of the harmonic average of the first moments of the structure factors.
	}
	\label{fig:SF}
\end{figure}

Figure~\ref{fig:SF} shows the scaled structure factor $S(kL_p)/L_p^3$ for all four cases, surface and bulk diffusion with $\left< \phi \right>=0.50$ and $\left< \phi \right>=0.36$, at the latest simulation time for each case.
As expected based on Porod's Law (Eq.\ \ref{eq:porodstruc}), the structure factors for all cases coincide at large values of the scaled wavenumber $kL_p$.
The characteristic wavelength $\lambda=1/k_0$, where $k_0$ is the first moment of $S(k)$, was obtained for each case in Fig.\ \ref{fig:SF}.
Although the height and width of the structure factor in the four cases differ, the scaled wavelengths $\lambda/L_p$ are very similar: at $\left<\phi\right>=0.50$, 2.32 for bulk diffusion and 2.31 for surface diffusion, and at $\left<\phi\right>=0.36$, 2.36 for bulk diffusion and 2.34 for surface diffusion.
The mean characteristic wavelength for all datasets is therefore $\bar \lambda/L_p = 2.33 \pm 0.02$, where the error is the standard deviation.
The scaled wavenumber corresponding to $\bar \lambda$, $L_p/\bar \lambda$, which is also the harmonic average of the first moments of $S(kL_p)$ for each case, is indicated by the green dash-dot line in Fig.~\ref{fig:SF}.

\begin{figure}
	\centering
	\includegraphics[width=16.3cm]{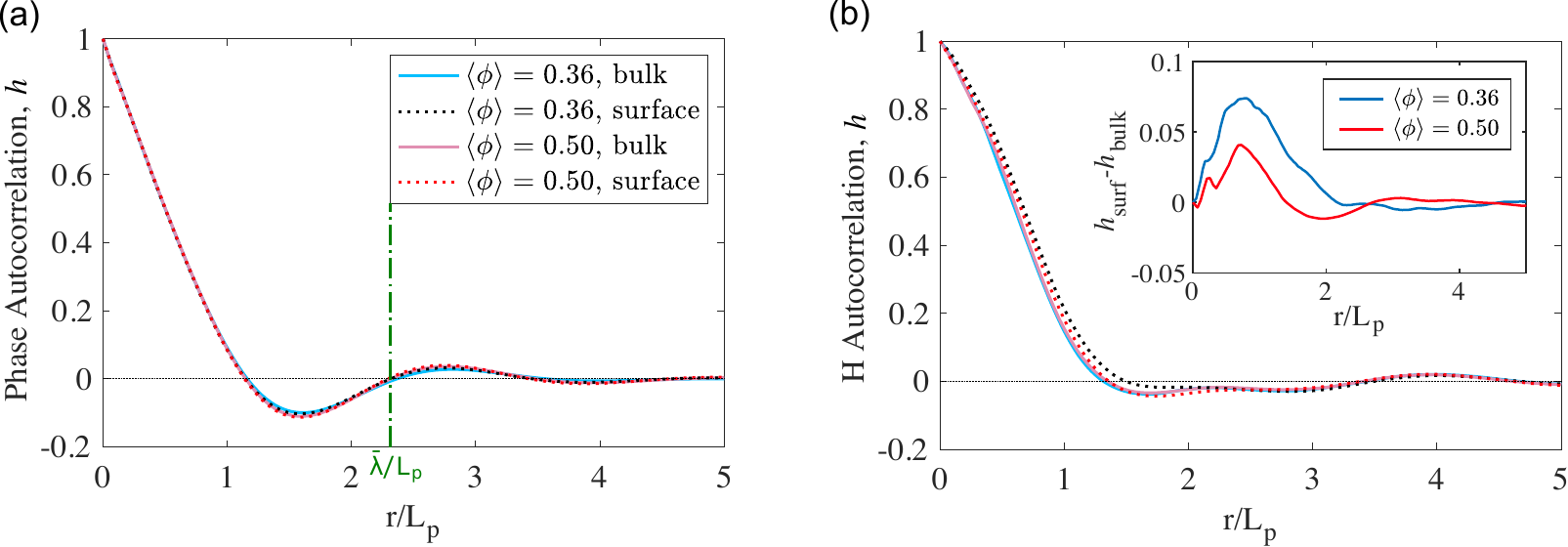}
	\caption{
	Plots of (a) Pearson phase autocorrelation and (b) Pearson $H$ autocorrelation for the $\left<\phi\right>=0.36$ structure with evolution governed by bulk diffusion (cyan line) and surface diffusion (black dashed line) and for the $\left<\phi\right>=0.50$ structure with evolution governed by bulk diffusion (light magenta line) and surface diffusion (red dashed line). The green dash-dot line in (a) represents the average characteristic wavelength for the four structures, i.e., $\bar{\lambda}/L_p$. The inset in (b) plots the difference between the Pearson $H$ autocorrelations of structures evolved with different coarsening mechanisms (i.e., $h_{surf}-h_{bulk}$) at the same nominal volume fractions, $\left<\phi\right>=0.36$ (blue line) and $\left<\phi\right>=0.50$ (red line).
	}
	\label{fig:autocorr_new}
\end{figure}

Figure~\ref{fig:autocorr_new}a shows the Pearson phase autocorrelation for all four cases at the latest simulation time for each case. 
(The time evolution of the autocorrelations is considered in the Supplemental Material \cite{supplemental}).
In each case, there is a peak at $h(0)=1$ from which the phase autocorrelations decrease with the slope given by Porod's Law (Eq.\ \ref{eq:porodlaw}), $h'(0)=-1/L_p$.
As $r/L_p$ increases, the phase autocorrelations decay below zero and then oscillate about zero with decreasing amplitude.
The wavelengths of these oscillations are related to the structure factor: the average scaled characteristic wavelength $\bar \lambda/L_p$ computed from the structure factors is indicated in Fig.~\ref{fig:autocorr_new}a by a vertical green dash-dot line, where it coincides with the second zero of the autocorrelations.
As shown in Ref.~\cite{Sun2018}, the distance between neighboring phases, the center-to-center distance between channels of the same phase, is approximately given by the location of the first local maximum of the phase autocorrelation.
This local maximum occurs around $r=2.8L_p$ for all cases.
The interphase distance (the space between the center of channels of opposite phase) is given by half the value of the first local maximum of the phase autocorrelation, $r=1.4L_p$.
Although these distances are the same for each structure, $L_p$ is different. 
Thus the differences between cases due to volume fraction are encapsulated in the scaling factor, $L_p = 4V_f(1-V_f)S_V^{-1}$.
No substantial differences due to coarsening mechanism are observed.
For the leveled-wave modeled structures, Soyarslan et al.~\cite{soyarslan20183d} find that the location of the first local maximum in the phase autocorrelation is always at a length $\alpha \lambda$, where $\lambda$ is the characteristic wavelength of the leveled-wave structure and $\alpha$ is a constant approximately equal to 1.23.
As the local maximum in Fig.~\ref{fig:autocorr_new}a occurs around $r=2.8L_p$ and $\bar{\lambda}/L_p$=2.33, the value of $\alpha$ is similar across both studies.

Figure~\ref{fig:autocorr_new}b displays the Pearson $H$ autocorrelation for all four cases.
For each structure, the $H$ autocorrelation is less oscillatory than the phase autocorrelation and shows a large region of anti-correlation.
However, locations of extrema are similar between the phase and $H$ autocorrelations, despite the extrema themselves differing in magnitude and sign.
Differences in $H$ autocorrelations due to volume fraction are largely captured by the scaling factor $L_p$ in the bulk-diffusion cases (as was observed for the phase autocorrelations in all cases), but the surface-diffusion cases do show an effect of volume fraction not captured by $L_p$.
The effect of coarsening mechanism on the $H$ autocorrelations is shown in the inset to Fig.\ \ref{fig:autocorr_new}b, which plots the difference between the $H$ autocorrelations for each mechanism, $h_{\mathrm{surf}}(r/L_p)$ and  $h_{\mathrm{bulk}}(r/L_p)$, at $\left<\phi\right>=0.36$ and $\left<\phi\right>=0.50$.
We omit the corresponding plot for phase autocorrelations because the differences between them are at most 0.01 between any two cases, which is comparable to the fluctuation of the phase autocorrelations between different times in their converged regimes (as determined from plots provided in the Supplemental Material \cite{supplemental}).
The $H$ autocorrelations have both larger differences due to coarsening mechanism and larger differences over the converged regime (at most 0.05 \cite{supplemental}), but the difference at $\left<\phi\right>=0.36$ appears to be significant, and we include the $\left<\phi\right>=0.50$ case for comparison.
The difference $h_{\mathrm{surf}} - h_{\mathrm{bulk}}$ at both volume fractions peaks near $r/L_p=0.7-0.8$, then decays to a negative value as $r/L_p$ increases, and finally converges toward zero at large $r/L_p$.
The initial peak value is approximately twice as large at $\left<\phi\right>=0.36$ as at $\left<\phi\right>=0.50$, indicating that the effect of coarsening mechanism is more pronounced when the volume fraction is away from the 50\% symmetric point.
Positive values of $h_{\mathrm{surf}} - h_{\mathrm{bulk}}$ at low $r/L_p$ indicate that curvatures in the structures are more spatially correlated when the dynamics are governed by surface diffusion, as the $H$ autocorrelation is larger for the surface-diffusion structures at the same value of $r/L_p$.
Thus, at $\left<\phi\right>=0.36$, points on the interface at small distances from each other are more likely to have the same mean curvature when the coarsening mechanism is surface diffusion as opposed to bulk diffusion.

\subsubsection{Scaled topological and morphological characteristics with comparisons to model structures \label{sec:stats}}
To summarize the morphological and topological differences between mechanisms and provide a basis for comparison to the model structures, we compute characteristic values of $\left< H/S_V\right>$, $\sigma_{H/S_V}$, and $g_V S_V^{-3}$ by averaging the values plotted in Fig.\ \ref{fig:evolution} over the converged regime for each case.
These characteristics are given in Tables \ref{tab:stats50} and \ref{tab:stats36} alongside values for the model structures.
Table \ref{tab:stats50} compares the structures at $V_f=0.50$ ($\left<\phi\right>=0.50$) and Table \ref{tab:stats36} compares the structures at $V_f=0.36$ ($\left<\phi\right>=0.36$).

As seen in Table \ref{tab:stats50}, the coarsened structures at $\left< \phi \right>=0.50$ are quantitatively similar; the surface-diffusion case has lower $\sigma_{H/S_V}$ and slightly lower $g_V S_V^{-3}$, with no difference in $\left< H/S_V\right>$.
$\left< H/S_V\right>=0$ is expected at this volume fraction due to the symmetry between the phases, and this is found in both the coarsened structures and the model structures.
The coarsened structures have values of $\sigma_{H/S_V}$ and $g_VS_V^{-3}$ between those of the model structures: the Schoen G surface overestimates $g_V S_V^{-3}$ and has $\sigma_{H/S_V}=0$, while the leveled-wave structure underestimates $g_V S_V^{-3}$ and overestimates $\sigma_{H/S_V}$.

\begin{table}
\centering
\caption{\label{tab:stats50} Scaled topology and morphological statistics for structures with $V_f=0.50$}
\begin{ruledtabular}
\begin{tabular}{l l l l l}
 & \multicolumn{2}{c}{Coarsened structures} & \multicolumn{2}{c}{Model structures} \\
 & surface & bulk & leveled-wave & Schoen G \\
\hline
$\left< H/S_V \right>$ & 0.00 & 0.00  & 0.00 & 0.00 \\
$\sigma_{H/S_V}$ & 0.30 & 0.32  & 0.50 & 0.00 \\
$g_VS_V^{-3}$ & 0.128 & 0.132 & 0.098 & 0.168  \\
\end{tabular}
\end{ruledtabular}
\end{table}

For the coarsened structures at $\left< \phi \right>=0.36$ in Table \ref{tab:stats36}, we find, as at $\left< \phi \right>=0.50$, that $g_V S_V^{-3}$ and $\sigma_{H/S_V}$ are lower for surface diffusion than for bulk diffusion, although the difference in $g_V S_V^{-3}$ is much larger than at $\left< \phi \right>=0.50$.
There is also a nonzero difference in $\left< H/S_V\right>$ between coarsened structures at $\left< \phi \right>=0.36$, with the surface-diffusion case having higher $\left< H/S_V\right>$ than the bulk-diffusion case.
The model structures at $V_f=0.36$ reasonably approximate the values of $\left< H/S_V\right>$ from the coarsened structures: the leveled-wave structure is closer, matching the bulk-diffusion structure, while the Schoen G-type surface has lower $\left< H/S_V\right>$ than either coarsened structure.
Still, the trend in the discrepancies in $\sigma_{H/S_V}$ and $g_V S_V^{-3}$ observed at $V_f=0.50$ are also found at $V_f=0.36$: the leveled-wave structure underestimates $g_V S_V^{-3}$ and overestimates $\sigma_{H/S_V}$, the Schoen G-type surface does the opposite, and thus the values of $g_V S_V^{-3}$ and $\sigma_{H/S_V}$ for the coarsened structures always lie in between those of the two model structures.

\begin{table}
\centering
\caption{\label{tab:stats36} Scaled topology and morphological statistics for structures with $V_f=0.36$}
\begin{ruledtabular}
\begin{tabular}{l l l l l}
 & \multicolumn{2}{c}{Coarsened structures} & \multicolumn{2}{c}{Model structures} \\
 & surface & bulk & leveled-wave & Schoen G \\
\hline
$\left< H/S_V \right>$ & 0.41 & 0.38 & 0.38 & 0.35 \\
$\sigma_{H/S_V}$ & 0.36 & 0.37  & 0.57 & 0.00 \\
$g_VS_V^{-3}$ & 0.114 & 0.129 & 0.097 & 0.188  \\
\end{tabular}
\end{ruledtabular}
\end{table}

Comparing between the coarsened structures in Tables \ref{tab:stats50} and \ref{tab:stats36}, we find higher $\left<H/S_V\right>$, higher $\sigma_{H/S_V}$, and lower $g_V S_V^{-3}$ at $\left< \phi \right>=0.36$ compared to $\left< \phi \right>=0.50$.
Furthermore, the differences due to volume fraction are larger than the differences due to coarsening mechanism (except in the case of scaled genus density, which is very similar between the bulk-diffusion structures at $\left< \phi \right>=0.50$ and $\left< \phi \right>=0.36$).
The leveled-wave structure qualitatively predicts the differences in $g_V S_V^{-3}$ and $\sigma_{H/S_V}$ between volume fractions, and its predictions for $\left< H/S_V \right>$ are quantitatively accurate for the bulk-diffusion cases.
The Schoen G-type surfaces provide reasonable agreement for $\left< H/S_V \right>$ at both volume fractions, but they cannot predict $\sigma_{H/S_V}$ due to the nature of minimal surfaces, and their trend for $g_V S_V^{-3}$ is opposite of the coarsened structures ($g_V S_V^{-3}$ of the Schoen G-type surfaces increases from $V_f=0.50$ to $V_f=0.36$ because the surface area per unit cell decreases while genus remains constant).

The data in Tables \ref{tab:stats50} and \ref{tab:stats36} can be compared to the results of previous simulation studies.
Our results for coarsening via bulk diffusion are generally in agreement with those of Kwon et al.\ \cite{Kwon2007,Kwon2010, Kwon2007thesis}, who report $g_V S_V^{-3}=0.134\pm 0.005$ for both $\left< \phi \right>=0.50$ and $\left< \phi \right>=0.36$, $\sigma_{H/S_V}=0.34$ at $\left< \phi \right>=0.50$, and $\left<H/S_V\right>=0.39$ at $\left< \phi \right>=0.36$.
For coarsening via surface diffusion, Geslin et al.\ \cite{Geslin2019} provide genus densities and statistics of mean curvature scaled by $\lambda$ rather than $S_V^{-1}$.
Since we have computed both $\lambda$ and $S_V^{-1}$ for structures comparable to theirs, we are able convert between these scalings, and details of these calculations are provided in the Supplemental Material \cite{supplemental}.
For scaled genus density, the values from Geslin et al.\ convert to $g_V S_V^{-3}=0.122$ at $\left< \phi \right>=0.50$ and $g_V S_V^{-3}=0.108$ at $\left< \phi \right>=0.375$, which are in reasonable agreement with our results in Tables \ref{tab:stats50} and \ref{tab:stats36}.
Geslin et al.\ find slightly lower values for $\sigma_{H/S_V}$, with $\sigma_{H/S_V}=0.27$ at $\left< \phi \right>=0.50$ and $\sigma_{H/S_V}=0.31$ at $\left< \phi \right>=0.375$.
This may constitute reasonable agreement given the sensitivity of $\sigma_{H/S_V}$ to extreme values of $H$, and the trend found in $\sigma_{H/S_V}$ by Geslin et al.\ with respect to volume fraction  matches our results.
While the values of $g_V S_V^{-3}$ reported here for coarsening via surface diffusion are consistent with those of Geslin et al., they appear to disagree with the results of a kinetic Monte Carlo study by Li et al.\ \cite{li2019topology}, which found agreement between coarsened structures and the leveled-wave model structure within the range of volume fractions considered here.

\subsubsection{Interfacial shape distributions}
\begin{figure*}
	\centering
	\includegraphics{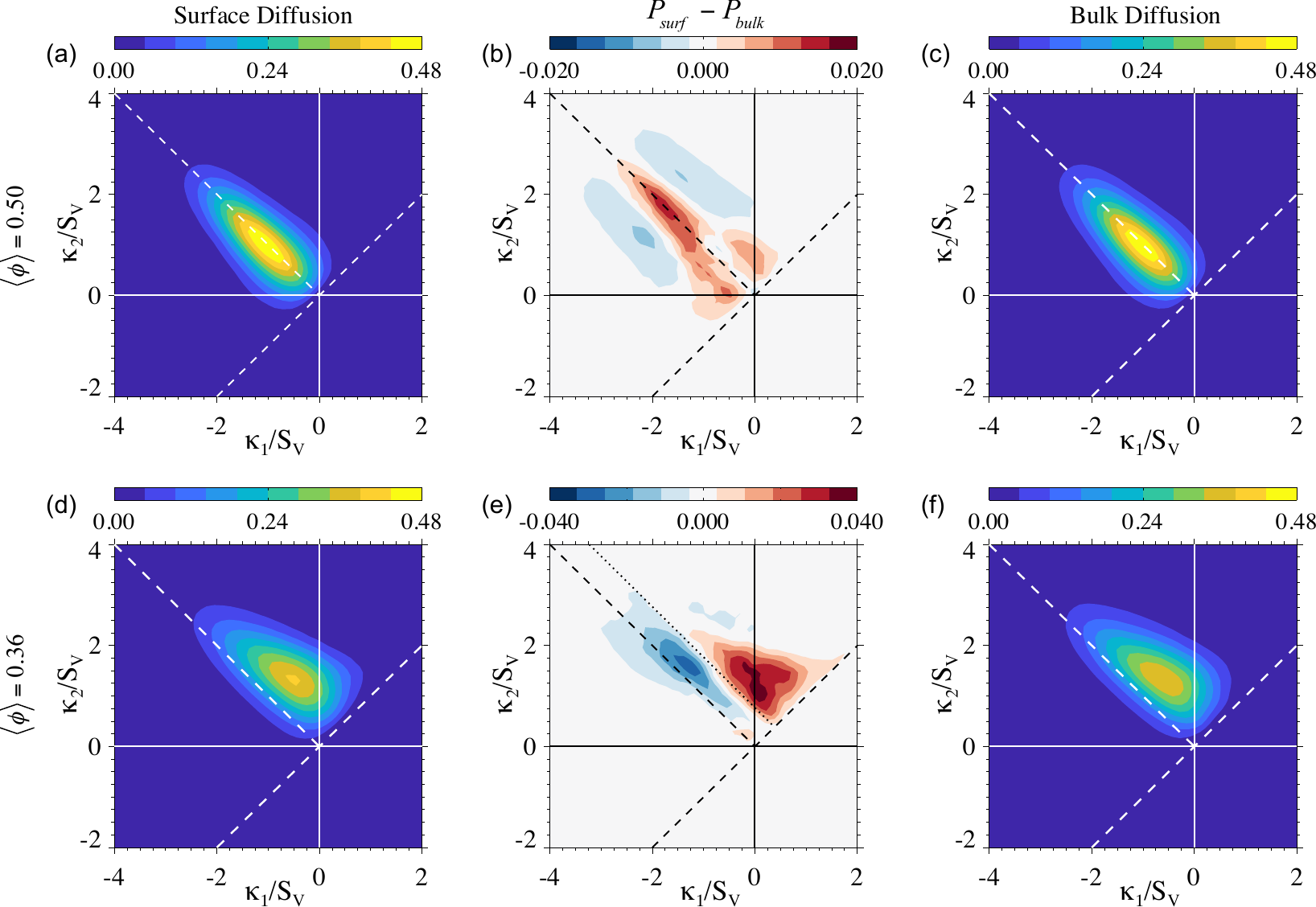}
	\caption{
		Comparison of converged average ISDs at (a-c) $\left< \phi \right> =0.50$ and (d-f) $\left< \phi \right> =0.36$, showing (a,d) the surface-diffusion ISDs, (b,e) the difference between bulk- and surface-diffusion ISDs (note the change in scale from (b) to (e)) , $P_{surf} - P_{bulk}$, and (c,f) the bulk-diffusion ISDs.  The diagonal dashed lines indicate $\kappa_1=-\kappa_2$ ($H=0$) and $\kappa_1=\kappa_2$ (spherically shaped interfaces).  The dotted line in (e) at $H/S_V=0.38$ indicates the average scaled mean curvature of the bulk-diffusion case (f).
	}
	\label{fig:ISDs}
\end{figure*}

The converged average ISDs $\bar P(\kappa_1/S_V,\kappa_2/S_V)$ introduced in Section \ref{sec:evolution} are shown in Fig.\ \ref{fig:ISDs}.
Figures \ref{fig:ISDs}a-c depict the $\left<\phi\right>=0.50$ cases, while Figs.\ \ref{fig:ISDs}d-f depict the $\left<\phi\right>=0.36$ cases.
The surface-diffusion ISDs are given in Figs.\ \ref{fig:ISDs}a and \ref{fig:ISDs}d, the bulk-diffusion ISDs are given in Figs.\ \ref{fig:ISDs}c and \ref{fig:ISDs}f, and Figs.\ \ref{fig:ISDs}b and \ref{fig:ISDs}e show the differences between them.
The line $\kappa_1 = -\kappa_2$ corresponds to zero mean curvature ($H = 0$), and is indicated on the ISDs by a dashed line extending from the origin.
The lines $\kappa_1 = 0$ and $\kappa_2 =0$ correspond to zero Gaussian curvature ($K=0$), where the interfacial shape corresponds to that of a cylinder.

Hyperbolic interfaces lie in the quadrant $\kappa_1 < 0$, $\kappa_2 > 0$, and they comprise the majority of interfaces in all four ISDs.
The differences between ISDs solely due to volume fraction are relatively large, with the normed difference $||P_{0.50}-P_{0.36}||_1$ equal to 0.99 for surface diffusion and 0.94 for bulk diffusion.
The $\left< \phi \right> =0.50$ ISDs are concentrated closely around the $H=0$ line, while the $\left< \phi \right> =0.36$ ISDs are shifted toward the upper right, indicating larger populations of interfaces with larger $H$.
This is consistent with the differences in $\left<H/S_V\right>$ between Tables \ref{tab:stats50} and \ref{tab:stats36} and the differences in $H/S_V$ between structures in Fig.\ \ref{fig:structures}, where the $\left< \phi \right> =0.36$ structure has substantially more areas with positive mean curvature than the $\left< \phi \right> =0.50$ structure.
Additionally, the $\left< \phi \right> =0.36$ ISDs are more diffuse, which is consistent with their higher values of $\sigma_{H/S_V}$.

At $\left< \phi \right> =0.50$, the difference between ISDs due to coarsening mechanism (Fig.\ \ref{fig:ISDs}b) is very small.
Quantitatively, the normed difference $||P_{surf}-P_{bulk}||_1$ between the ISDs is $0.04$, only slightly larger than variations in the ISDs over time, which are in the range $0.01-0.03$ \cite{supplemental}.
The surface-diffusion ISD has more low-curvature area near the origin, including more elliptic area, and its higher-curvature areas are more closely concentrated about the $H=0$ line than in the bulk-diffusion ISD, which may result in its lower $\sigma_{H/S_V}$ in Table \ref{tab:stats50}.
The difference between surface- and bulk-diffusion ISDs at $\left< \phi \right> =0.36$ is shown in Fig.\ \ref{fig:ISDs}e.
Quantitatively, $||P_{surf}-P_{bulk}||_1 = 0.09$, which is substantially larger than the variation in time of either ISD ($0.02-0.04$ \cite{supplemental}).
The line $H/S_V=0.38$, representing $\left< H/S_V\right>$ of the bulk-diffusion case, has been drawn onto Fig.\ \ref{fig:ISDs}e as a dotted line.
The surface-diffusion ISD contains more area to the upper right of this line (with higher $H/S_V$) and less area to the lower left of it (with lower $H/S_V$).
This decrease in area at lower $H/S_V$ corresponds primarily to hyperbolic interfaces with highly negative $K$, such as the necks observed in Fig.\ \ref{fig:structures}.
Much of the increase in area at higher $H/S_V$ corresponds to an increase in elliptic areas with $\kappa_1  > 0$ (and thus $K>0$ and $H>0$), which can be attributed to the caps that form after necks pinch off.
Both the decrease in hyperbolic area and the increase in elliptic area are expected to reduce $g_V S_V^{-3}$ in the surface-diffusion structure because of the Gauss-Bonnet theorem (Eq.\ \ref{eq:GaussBonnetsc}), matching the trend in Table \ref{tab:stats36}.

\subsubsection{Discussion}
Coarsening via surface diffusion at nominal volume fractions of 36\% and 50\% (system compositions of $\left<\phi\right>=0.36$ and $\left<\phi\right>=0.50$, respectively) resulted in self-similarly evolving structures that are quite close to those evolving via bulk diffusion.
However,  quantitative characterization of the structures revealed differences in morphology and topology due to the underlying coarsening mechanism.
The surface-diffusion structures have lower scaled genus density and lower standard deviation of scaled mean curvature.
At $\left<\phi\right>=0.36$, the surface diffusion structure has lower short-range spatial variation in $H$ than the bulk diffusion structure, as assessed by its autocorrelation.
The ISDs for surface diffusion contained relatively more elliptic area than the bulk-diffusion ISDs, which is consistent with their lower scaled genus density.

The differences in morphology and topology between the surface- and bulk-diffusion cases are surprisingly subtle given the difference in dynamics between them.
To fully understand why this is the case, we would need to develop a better theoretical connection between coarsening dynamics and morphology.
One approach is to examine the topological singularities that occur during coarsening \cite{Bouttes2016,andrews2020simulation}, specifically the pinching-off of necks and the flattening of the resulting cones into rounded caps \cite{Bernoff1998,Wong1998,Gorshkov2017,Lai2019,Aagesen2010,Aagesen2011}.
Since all of the structures have a substantial genus density and a negligible number of particles, the pinching-off of necks is the dominant type of topological singularity in our structures.
Since this process results in hyperbolic interfaces (necks) turning into elliptic areas (caps), the higher fraction of elliptic area in the surface diffusion structures (and the lower scaled genus density) may represent slower evolution of caps compared to necks relative to the bulk-diffusion case.
Caps typically flatten rapidly due to their high mean curvature relative to the surrounding structure, and it seems reasonable that restricting transport to the surface could slow this flattening process disproportionately to the pinching-off process.
However, while pinching-off of necks has been analyzed for surface diffusion \cite{Bernoff1998,Wong1998,Gorshkov2017,Lai2019}, and for bulk diffusion when the diffusivities of the phases are highly dissimilar \cite{Aagesen2010,Aagesen2011}, it has not yet been examined for bulk diffusion with the same diffusivity in both phases.
Such investigation is needed in order to fully assess the role of topological singularities on morphological evolution during coarsening.

In addition to the overall effects of coarsening mechanism on morphology, we have found an interestingly strong interaction between coarsening mechanism and volume fraction: differences in morphology between surface and bulk diffusion are larger at $\left<\phi\right>=0.36$ than at $\left<\phi\right>=0.50$.
We observe this trend in several microstructural characteristics: ISDs, $H$ autocorrelations, scaled genus densities, and average scaled mean curvature.
Only the standard deviation of scaled mean curvature provides a possible counterexample, with a slightly larger difference between coarsening mechanisms at $\left<\phi\right>=0.50$ than $\left<\phi\right>=0.36$.
To explain this trend, we examine differences between coarsening mechanisms that might result in a volume fraction dependence in the bulk diffusion case but not in the surface diffusion case.
While surface diffusion only allows diffusion between neighboring points on the interface, bulk diffusion allows interfaces to interact nonlocally.
In the latter, the local chemical potential at a point on the interface can affect the local interfacial velocity far from that point.
To approximate these interactions, a uniform mean-field chemical potential is often introduced.
In bicontinuous structures, this potential can be reasonably assumed to be proportional to $\left< H \right>$ \cite{Fife2014, andrews2020simulation} (see also the Supplemental Material \cite{supplemental}).
Since $\left< H /S_V \right>$ changes considerably between $\left<\phi\right>=0.50$ and $\left<\phi\right>=0.36$ (from zero to 0.38-0.41), the effect of a mean-field chemical potential is one possible explanation for the greater difference between coarsening mechanisms at $\left<\phi\right>=0.36$ compared to $\left<\phi\right>=0.50$.
While our study focused on comparisons between coarsening mechanisms at the same volume fraction, we also note that the differences in self-similar morphology and topology due to changes in nominal volume fraction are smaller for bulk diffusion than for surface diffusion.
This effect is most apparent in $g_VS_V^{-3}$ and the $H$ autocorrelations, but it is also found to a lesser extent in the ISDs, $\left< H /S_V \right>$, and $\sigma_{H/S_V}$.
Thus, the mean-field chemical potential in the bulk-diffusion case may have a stabilizing effect on the morphology and topology of bulk-diffusion structures as volume fraction decreases.

Although the model structures considered above were constructed solely on the basis of mathematical descriptions with no underlying physics of coarsening, they provide valuable insight into the effect of volume fraction on its own.
Both the Schoen G family of surfaces and the leveled-wave structures approximately predict the change in $\left< H /S_V \right>$ due to volume fraction, with quantitative agreement between the leveled-wave and bulk-diffusion structures.
In $g_VS_V^{-3}$ and $\sigma_{H/S_V}$, the leveled-wave structures also have the same qualitative trends with respect to volume fraction as the coarsened structures.
Where the model structures fail, they do so in ways that could perhaps be expected based on their construction.
The Schoen G family of surfaces necessarily have higher $g_VS_V^{-3}$ than the coarsened structures because they lack the caps of elliptic area that result from pinching-off events during coarsening.
The higher $\sigma_{H/S_V}$ and lower $g_VS_V^{-3}$ of the leveled-wave structures compared to the coarsened structures might indicate that the leveled-wave structures have more or higher-curvature elliptic area, which would flatten out during coarsening, which does not occur with the leveled-wave structure. 

\section{Conclusions}
We have conducted three-dimensional phase field simulations of phase separation and coarsening via two mechanisms, surface diffusion and bulk diffusion, at nominal volume fractions of $36\%$ and $50\%$ (average concentrations of $\left< \phi\right>=0.36$ and $\left< \phi\right>=0.50$).
All of the simulations resulted in bicontinuous microstructures, with no independent particles at $50\%$ and a negligible volume and number of particles at $36\%$.
Scaled genus density was used to define convergence to self-similar coarsening for each structure, an approach that was validated by examining convergence of interfacial shape distributions (ISDs) and the autocorrelations of phase and mean curvature.
Kinetics of coarsening via surface and bulk diffusion were found to agree with their respective power laws for self-similar evolution,  $S_V^{-1} \propto t^{1/4}$ and $S_V^{-1} \propto t^{1/3}$.

To analyze how coarsening mechanism affects morphology, we characterized and compared the self-similar morphologies for each coarsening mechanism and volume fraction.
Average scaled mean curvature $\left<H/S_V\right>$, standard deviation of scaled mean curvature $\sigma_{H/S_V}$, and scaled genus density $g_V S_V^{-3}$ were averaged over the self-similar regime for each structure, and the resulting values were found to agree reasonably with those in literature (Refs.\ \cite{Kwon2010,Geslin2019}).
ISDs, structure factors, and Pearson autocorrelations of phase and interfacial mean curvature were calculated for the self-similar regime and compared between coarsening mechanisms.
Overall, the structures coarsening by surface diffusion are remarkably similar to those coarsening by bulk diffusion, despite the different kinetics.
However, a close examination of structures reveals that  structures coarsening via surface diffusion had slightly lower scaled genus densities and standard deviations of scaled mean curvature, slightly more elliptic area (area with positive Gaussian curvature) in their ISDs, and, at 36\% volume fraction, greater spatial correlations of mean curvatures at relatively short distances.
Phase autocorrelations and structure factors were also very similar across all volume fractions and coarsening mechanisms when distance was scaled by $L_p=4V_f(1-V_f)S_V^{-1}$, a factor based on Porod's law.
In most other metrics, there was a strong interaction between the effects of volume fraction and coarsening mechanism, with much larger differences at $36\%$ than at $50\%$.
We have hypothesized possible explanations for the effect of coarsening mechanism on morphology (and its interaction with the effect of volume fraction), but more rigorous theoretical development is needed before they are fully understood.

In addition to comparing between coarsening mechanisms, we compared the self-similar coarsened morphologies to two model structures for bicontinuous structures: the family of constant-mean-curvature surfaces based on the Schoen G triply periodic minimal surface \cite{Jung2007minimal} and the random leveled-wave structures that result from thresholding a field of superposed sinusoids with random phase and orientation \cite{Teubner1991random,soyarslan20183d}.
These model structures cannot account for any effects of coarsening mechanism, but they do predict how morphology might change as a function of volume fraction.
Both model structures provide reasonable predictions for $\left<H/S_V\right>$, but the values of $\sigma_{H/S_V}$ and $g_V S_V^{-3}$ that they predict differ substantially from those of the coarsened structures.
The values of $\sigma_{H/S_V}$ and $g_V$ for the coarsened structures were always between those of the two model structures, with the leveled-wave structures having higher $\sigma_{H/S_V}$ and lower $g_V S_V^{-3}$ than both the coarsened structures and the Schoen G family of surfaces.
Documentation of these differences will be useful to researchers attempting to model the properties of bicontinuous structures that have undergone coarsening.

\section*{Data Availability}
Analysis results presented in this work and selected raw simulation output have been released on Materials Commons (materialscommons.org) and are available at the following doi's: 10.13011/m3-w4pw-sy79 (coarsening via bulk diffusion at 50\% volume fraction), 10.13011/m3-gnvz-b725 (coarsening via bulk diffusion at 36\%), 10.13011/m3-3591-7648 (coarsening via surface diffusion at 36\% and 50\%), and  10.13011/m3-xjd4-bm50 (coarsening via the alternative surface diffusion model discussed in the SM \cite{supplemental}).

\section*{Acknowledgements}
This work was supported by United States Department of Energy grants DE-SC0015394 and DE-FG02-99ER45782, and by the Natural Sciences and Engineering Research Council of Canada (NSERC) [PGSD3-516809-2018].
Computational resources for simulations were provided by the Extreme Science and Engineering Discovery Environment (XSEDE), which is supported by the National Science Foundation grant number OCI-1053575, under allocation No.\ TG-DMR110007, and additional resources for code testing and data analysis were provided by Advanced Research Computing (ARC) at the University of Michigan.


\begin{thebibliography}{62}%
\makeatletter
\providecommand \@ifxundefined [1]{%
 \@ifx{#1\undefined}
}%
\providecommand \@ifnum [1]{%
 \ifnum #1\expandafter \@firstoftwo
 \else \expandafter \@secondoftwo
 \fi
}%
\providecommand \@ifx [1]{%
 \ifx #1\expandafter \@firstoftwo
 \else \expandafter \@secondoftwo
 \fi
}%
\providecommand \natexlab [1]{#1}%
\providecommand \enquote  [1]{``#1''}%
\providecommand \bibnamefont  [1]{#1}%
\providecommand \bibfnamefont [1]{#1}%
\providecommand \citenamefont [1]{#1}%
\providecommand \href@noop [0]{\@secondoftwo}%
\providecommand \href [0]{\begingroup \@sanitize@url \@href}%
\providecommand \@href[1]{\@@startlink{#1}\@@href}%
\providecommand \@@href[1]{\endgroup#1\@@endlink}%
\providecommand \@sanitize@url [0]{\catcode `\\12\catcode `\$12\catcode
  `\&12\catcode `\#12\catcode `\^12\catcode `\_12\catcode `\%12\relax}%
\providecommand \@@startlink[1]{}%
\providecommand \@@endlink[0]{}%
\providecommand \url  [0]{\begingroup\@sanitize@url \@url }%
\providecommand \@url [1]{\endgroup\@href {#1}{\urlprefix }}%
\providecommand \urlprefix  [0]{URL }%
\providecommand \Eprint [0]{\href }%
\providecommand \doibase [0]{https://doi.org/}%
\providecommand \selectlanguage [0]{\@gobble}%
\providecommand \bibinfo  [0]{\@secondoftwo}%
\providecommand \bibfield  [0]{\@secondoftwo}%
\providecommand \translation [1]{[#1]}%
\providecommand \BibitemOpen [0]{}%
\providecommand \bibitemStop [0]{}%
\providecommand \bibitemNoStop [0]{.\EOS\space}%
\providecommand \EOS [0]{\spacefactor3000\relax}%
\providecommand \BibitemShut  [1]{\csname bibitem#1\endcsname}%
\let\auto@bib@innerbib\@empty
\bibitem [{\citenamefont {Bray}(2002)}]{Bray2002}%
  \BibitemOpen
  \bibfield  {author} {\bibinfo {author} {\bibfnamefont {A.~J.}\ \bibnamefont
  {Bray}},\ }\bibfield  {title} {\bibinfo {title} {Theory of phase-ordering
  kinetics},\ }\href {https://doi.org/10.1080/00018730110117433} {\bibfield
  {journal} {\bibinfo  {journal} {Advances in Physics}\ }\textbf {\bibinfo
  {volume} {51}},\ \bibinfo {pages} {481} (\bibinfo {year} {2002})}\BibitemShut
  {NoStop}%
\bibitem [{\citenamefont {Herring}(1950)}]{Herring1950}%
  \BibitemOpen
  \bibfield  {author} {\bibinfo {author} {\bibfnamefont {C.}~\bibnamefont
  {Herring}},\ }\bibfield  {title} {\bibinfo {title} {Effect of change of scale
  on sintering phenomena},\ }\href {https://doi.org/10.1063/1.1699658}
  {\bibfield  {journal} {\bibinfo  {journal} {Journal of Applied Physics}\
  }\textbf {\bibinfo {volume} {21}},\ \bibinfo {pages} {301} (\bibinfo {year}
  {1950})}\BibitemShut {NoStop}%
\bibitem [{\citenamefont {Geslin}\ \emph {et~al.}(2019)\citenamefont {Geslin},
  \citenamefont {Buchet}, \citenamefont {Wada},\ and\ \citenamefont
  {Kato}}]{Geslin2019}%
  \BibitemOpen
  \bibfield  {author} {\bibinfo {author} {\bibfnamefont {P.-A.}\ \bibnamefont
  {Geslin}}, \bibinfo {author} {\bibfnamefont {M.}~\bibnamefont {Buchet}},
  \bibinfo {author} {\bibfnamefont {T.}~\bibnamefont {Wada}}, and\ \bibinfo
  {author} {\bibfnamefont {H.}~\bibnamefont {Kato}},\ }\bibfield  {title}
  {\bibinfo {title} {Phase-field investigation of the coarsening of porous
  structures by surface diffusion},\ }\href
  {https://doi.org/10.1103/PhysRevMaterials.3.083401} {\bibfield  {journal}
  {\bibinfo  {journal} {Phys. Rev. Materials}\ }\textbf {\bibinfo {volume}
  {3}},\ \bibinfo {pages} {083401} (\bibinfo {year} {2019})}\BibitemShut
  {NoStop}%
\bibitem [{\citenamefont {Kwon}\ \emph {et~al.}(2010)\citenamefont {Kwon},
  \citenamefont {Thornton},\ and\ \citenamefont {Voorhees}}]{Kwon2010}%
  \BibitemOpen
  \bibfield  {author} {\bibinfo {author} {\bibfnamefont {Y.}~\bibnamefont
  {Kwon}}, \bibinfo {author} {\bibfnamefont {K.}~\bibnamefont {Thornton}}, and\
  \bibinfo {author} {\bibfnamefont {P.}~\bibnamefont {Voorhees}},\ }\bibfield
  {title} {\bibinfo {title} {Morphology and topology in coarsening of domains
  via non-conserved and conserved dynamics},\ }\href
  {https://doi.org/10.1080/14786430903260701} {\bibfield  {journal} {\bibinfo
  {journal} {Philosophical Magazine}\ }\textbf {\bibinfo {volume} {90}},\
  \bibinfo {pages} {317} (\bibinfo {year} {2010})}\BibitemShut {NoStop}%
\bibitem [{\citenamefont {Henry}\ and\ \citenamefont
  {Tegze}(2018)}]{Henry2018}%
  \BibitemOpen
  \bibfield  {author} {\bibinfo {author} {\bibfnamefont {H.}~\bibnamefont
  {Henry}}and\ \bibinfo {author} {\bibfnamefont {G.}~\bibnamefont {Tegze}},\
  }\bibfield  {title} {\bibinfo {title} {Self-similarity and coarsening rate of
  a convecting bicontinuous phase separating mixture: Effect of the viscosity
  contrast},\ }\href {https://doi.org/10.1103/PhysRevFluids.3.074306}
  {\bibfield  {journal} {\bibinfo  {journal} {Phys. Rev. Fluids}\ }\textbf
  {\bibinfo {volume} {3}},\ \bibinfo {pages} {074306} (\bibinfo {year}
  {2018})}\BibitemShut {NoStop}%
\bibitem [{\citenamefont {Lifshitz}\ and\ \citenamefont
  {Slyozov}(1961)}]{Lifshitz1961}%
  \BibitemOpen
  \bibfield  {author} {\bibinfo {author} {\bibfnamefont {I.~M.}\ \bibnamefont
  {Lifshitz}}and\ \bibinfo {author} {\bibfnamefont {V.~V.}\ \bibnamefont
  {Slyozov}},\ }\bibfield  {title} {\bibinfo {title} {The kinetics of
  precipitation from supersaturated solid solutions},\ }\href
  {https://doi.org/10.1016/0022-3697(61)90054-3} {\bibfield  {journal}
  {\bibinfo  {journal} {Journal of Physics and Chemistry of Solids}\ }\textbf
  {\bibinfo {volume} {19}},\ \bibinfo {pages} {35} (\bibinfo {year}
  {1961})}\BibitemShut {NoStop}%
\bibitem [{\citenamefont {Wagner}(1961)}]{Wagner1961}%
  \BibitemOpen
  \bibfield  {author} {\bibinfo {author} {\bibfnamefont {C.}~\bibnamefont
  {Wagner}},\ }\bibfield  {title} {\bibinfo {title} {Theorie der {Alterung} von
  {Niederschl\"agen} durch {Uml\"osen} {(Ostwald-Reifung)}},\ }\href
  {https://doi.org/10.1002/bbpc.19610650704} {\bibfield  {journal} {\bibinfo
  {journal} {Z. Elektrochem}\ }\textbf {\bibinfo {volume} {65}},\ \bibinfo
  {pages} {581} (\bibinfo {year} {1961})}\BibitemShut {NoStop}%
\bibitem [{\citenamefont {Baldan}(2002)}]{Baldan2002}%
  \BibitemOpen
  \bibfield  {author} {\bibinfo {author} {\bibfnamefont {A.}~\bibnamefont
  {Baldan}},\ }\bibfield  {title} {\bibinfo {title} {Review progress in
  {Ostwald} ripening theories and their applications to nickel-base superalloys
  {Part} {I:} {Ostwald} ripening theories},\ }\href
  {https://doi.org/10.1023/A:1015388912729} {\bibfield  {journal} {\bibinfo
  {journal} {Journal of Materials Science}\ }\textbf {\bibinfo {volume} {37}},\
  \bibinfo {pages} {2171} (\bibinfo {year} {2002})}\BibitemShut {NoStop}%
\bibitem [{\citenamefont {Jinnai}\ \emph {et~al.}(2000)\citenamefont {Jinnai},
  \citenamefont {Nishikawa}, \citenamefont {Morimoto}, \citenamefont {Koga},\
  and\ \citenamefont {Hashimoto}}]{Jinnai2000}%
  \BibitemOpen
  \bibfield  {author} {\bibinfo {author} {\bibfnamefont {H.}~\bibnamefont
  {Jinnai}}, \bibinfo {author} {\bibfnamefont {Y.}~\bibnamefont {Nishikawa}},
  \bibinfo {author} {\bibfnamefont {H.}~\bibnamefont {Morimoto}}, \bibinfo
  {author} {\bibfnamefont {T.}~\bibnamefont {Koga}}, and\ \bibinfo {author}
  {\bibfnamefont {T.}~\bibnamefont {Hashimoto}},\ }\bibfield  {title} {\bibinfo
  {title} {Geometrical properties and interface dynamics: time evolution of
  spinodal interface in a binary polymer mixture at the critical composition},\
  }\href {https://doi.org/10.1021/la991024q} {\bibfield  {journal} {\bibinfo
  {journal} {Langmuir}\ }\textbf {\bibinfo {volume} {16}},\ \bibinfo {pages}
  {4380} (\bibinfo {year} {2000})}\BibitemShut {NoStop}%
\bibitem [{\citenamefont {McCue}\ \emph {et~al.}(2016)\citenamefont {McCue},
  \citenamefont {Benn}, \citenamefont {Gaskey},\ and\ \citenamefont
  {Erlebacher}}]{mccue2016dealloying}%
  \BibitemOpen
  \bibfield  {author} {\bibinfo {author} {\bibfnamefont {I.}~\bibnamefont
  {McCue}}, \bibinfo {author} {\bibfnamefont {E.}~\bibnamefont {Benn}},
  \bibinfo {author} {\bibfnamefont {B.}~\bibnamefont {Gaskey}}, and\ \bibinfo
  {author} {\bibfnamefont {J.}~\bibnamefont {Erlebacher}},\ }\bibfield  {title}
  {\bibinfo {title} {Dealloying and dealloyed materials},\ }\href@noop {}
  {\bibfield  {journal} {\bibinfo  {journal} {Annual Review of Materials
  Research}\ }\textbf {\bibinfo {volume} {46}},\ \bibinfo {pages} {263}
  (\bibinfo {year} {2016})}\BibitemShut {NoStop}%
\bibitem [{\citenamefont {Mendoza}\ \emph {et~al.}(2003)\citenamefont
  {Mendoza}, \citenamefont {Alkemper},\ and\ \citenamefont
  {Voorhees}}]{Mendoza2003}%
  \BibitemOpen
  \bibfield  {author} {\bibinfo {author} {\bibfnamefont {R.}~\bibnamefont
  {Mendoza}}, \bibinfo {author} {\bibfnamefont {J.}~\bibnamefont {Alkemper}},
  and\ \bibinfo {author} {\bibfnamefont {P.~W.}\ \bibnamefont {Voorhees}},\
  }\bibfield  {title} {\bibinfo {title} {The morphological evolution of
  dendritic microstructures during coarsening},\ }\href
  {https://doi.org/10.1007/s11661-003-0084-2} {\bibfield  {journal} {\bibinfo
  {journal} {Metallurgical and Materials Transactions A}\ }\textbf {\bibinfo
  {volume} {34}},\ \bibinfo {pages} {481} (\bibinfo {year} {2003})}\BibitemShut
  {NoStop}%
\bibitem [{\citenamefont {Kwon}\ \emph {et~al.}(2007)\citenamefont {Kwon},
  \citenamefont {Thornton},\ and\ \citenamefont {Voorhees}}]{Kwon2007}%
  \BibitemOpen
  \bibfield  {author} {\bibinfo {author} {\bibfnamefont {Y.}~\bibnamefont
  {Kwon}}, \bibinfo {author} {\bibfnamefont {K.}~\bibnamefont {Thornton}}, and\
  \bibinfo {author} {\bibfnamefont {P.~W.}\ \bibnamefont {Voorhees}},\
  }\bibfield  {title} {\bibinfo {title} {Coarsening of bicontinuous structures
  via nonconserved and conserved dynamics},\ }\href
  {https://doi.org/10.1103/PhysRevE.75.021120} {\bibfield  {journal} {\bibinfo
  {journal} {Phys. Rev. E}\ }\textbf {\bibinfo {volume} {75}},\ \bibinfo
  {pages} {021120} (\bibinfo {year} {2007})}\BibitemShut {NoStop}%
\bibitem [{\citenamefont {Kwon}(2007)}]{Kwon2007thesis}%
  \BibitemOpen
  \bibfield  {author} {\bibinfo {author} {\bibfnamefont {Y.}~\bibnamefont
  {Kwon}},\ }\emph {\bibinfo {title} {Morphology and Topology of Interfaces
  During Coarsening Via Nonconserved and Conserved Dynamics}},\ \href@noop {}
  {Ph.D. thesis},\ \bibinfo  {school} {Northwestern University} (\bibinfo
  {year} {2007})\BibitemShut {NoStop}%
\bibitem [{\citenamefont {Li}\ \emph {et~al.}(2019)\citenamefont {Li},
  \citenamefont {Dinh~Ng\^o}, \citenamefont {Markmann},\ and\ \citenamefont
  {Weissm\"uller}}]{li2019topology}%
  \BibitemOpen
  \bibfield  {author} {\bibinfo {author} {\bibfnamefont {Y.}~\bibnamefont
  {Li}}, \bibinfo {author} {\bibfnamefont {B.-N.}\ \bibnamefont {Dinh~Ng\^o}},
  \bibinfo {author} {\bibfnamefont {J.}~\bibnamefont {Markmann}}, and\ \bibinfo
  {author} {\bibfnamefont {J.}~\bibnamefont {Weissm\"uller}},\ }\bibfield
  {title} {\bibinfo {title} {Topology evolution during coarsening of nanoscale
  metal network structures},\ }\href@noop {} {\bibfield  {journal} {\bibinfo
  {journal} {Physical review materials}\ }\textbf {\bibinfo {volume} {3}},\
  \bibinfo {pages} {076001} (\bibinfo {year} {2019})}\BibitemShut {NoStop}%
\bibitem [{\citenamefont {G\'o\'{z}d\'{z}}\ and\ \citenamefont
  {Ho\l{}yst}(1996)}]{Gozdz1996minimal}%
  \BibitemOpen
  \bibfield  {author} {\bibinfo {author} {\bibfnamefont {W.~T.}\ \bibnamefont
  {G\'o\'{z}d\'{z}}}and\ \bibinfo {author} {\bibfnamefont {R.}~\bibnamefont
  {Ho\l{}yst}},\ }\bibfield  {title} {\bibinfo {title} {Triply periodic
  surfaces and multiply continuous structures from the {Landau} model of
  microemulsions},\ }\href {https://doi.org/10.1103/PhysRevE.54.5012}
  {\bibfield  {journal} {\bibinfo  {journal} {Phys. Rev. E}\ }\textbf {\bibinfo
  {volume} {54}},\ \bibinfo {pages} {5012} (\bibinfo {year}
  {1996})}\BibitemShut {NoStop}%
\bibitem [{\citenamefont {Jung}\ \emph {et~al.}(2007)\citenamefont {Jung},
  \citenamefont {Chu},\ and\ \citenamefont {Torquato}}]{Jung2007minimal}%
  \BibitemOpen
  \bibfield  {author} {\bibinfo {author} {\bibfnamefont {Y.}~\bibnamefont
  {Jung}}, \bibinfo {author} {\bibfnamefont {K.}~\bibnamefont {Chu}}, and\
  \bibinfo {author} {\bibfnamefont {S.}~\bibnamefont {Torquato}},\ }\bibfield
  {title} {\bibinfo {title} {A variational level set approach for surface area
  minimization of triply-periodic surfaces},\ }\href
  {https://doi.org/https://doi.org/10.1016/j.jcp.2006.10.007} {\bibfield
  {journal} {\bibinfo  {journal} {Journal of Computational Physics}\ }\textbf
  {\bibinfo {volume} {223}},\ \bibinfo {pages} {711 } (\bibinfo {year}
  {2007})}\BibitemShut {NoStop}%
\bibitem [{\citenamefont {Pia}\ \emph {et~al.}(2017)\citenamefont {Pia},
  \citenamefont {Brun}, \citenamefont {Aymerich},\ and\ \citenamefont
  {Delogu}}]{pia2017gyroidal}%
  \BibitemOpen
  \bibfield  {author} {\bibinfo {author} {\bibfnamefont {G.}~\bibnamefont
  {Pia}}, \bibinfo {author} {\bibfnamefont {M.}~\bibnamefont {Brun}}, \bibinfo
  {author} {\bibfnamefont {F.}~\bibnamefont {Aymerich}}, and\ \bibinfo {author}
  {\bibfnamefont {F.}~\bibnamefont {Delogu}},\ }\bibfield  {title} {\bibinfo
  {title} {Gyroidal structures as approximants to nanoporous metal foams: clues
  from mechanical properties},\ }\href@noop {} {\bibfield  {journal} {\bibinfo
  {journal} {Journal of Materials Science}\ }\textbf {\bibinfo {volume} {52}},\
  \bibinfo {pages} {1106} (\bibinfo {year} {2017})}\BibitemShut {NoStop}%
\bibitem [{\citenamefont {Cahn}(1965)}]{Cahn1965}%
  \BibitemOpen
  \bibfield  {author} {\bibinfo {author} {\bibfnamefont {J.~W.}\ \bibnamefont
  {Cahn}},\ }\bibfield  {title} {\bibinfo {title} {Phase separation by spinodal
  decomposition in isotropic systems},\ }\href
  {https://doi.org/10.1063/1.1695731} {\bibfield  {journal} {\bibinfo
  {journal} {The Journal of Chemical Physics}\ }\textbf {\bibinfo {volume}
  {42}},\ \bibinfo {pages} {93} (\bibinfo {year} {1965})},\ \Eprint
  {https://arxiv.org/abs/https://doi.org/10.1063/1.1695731}
  {https://doi.org/10.1063/1.1695731} \BibitemShut {NoStop}%
\bibitem [{\citenamefont {Teubner}(1991)}]{Teubner1991random}%
  \BibitemOpen
  \bibfield  {author} {\bibinfo {author} {\bibfnamefont {M.}~\bibnamefont
  {Teubner}},\ }\bibfield  {title} {\bibinfo {title} {Level surfaces of
  {Gaussian} random fields and microemulsions},\ }\href
  {https://doi.org/10.1209/0295-5075/14/5/003} {\bibfield  {journal} {\bibinfo
  {journal} {Europhysics Letters ({EPL})}\ }\textbf {\bibinfo {volume} {14}},\
  \bibinfo {pages} {403} (\bibinfo {year} {1991})}\BibitemShut {NoStop}%
\bibitem [{\citenamefont {Soyarslan}\ \emph {et~al.}(2018)\citenamefont
  {Soyarslan}, \citenamefont {Bargmann}, \citenamefont {Pradas},\ and\
  \citenamefont {Weissm{\"u}ller}}]{soyarslan20183d}%
  \BibitemOpen
  \bibfield  {author} {\bibinfo {author} {\bibfnamefont {C.}~\bibnamefont
  {Soyarslan}}, \bibinfo {author} {\bibfnamefont {S.}~\bibnamefont {Bargmann}},
  \bibinfo {author} {\bibfnamefont {M.}~\bibnamefont {Pradas}}, and\ \bibinfo
  {author} {\bibfnamefont {J.}~\bibnamefont {Weissm{\"u}ller}},\ }\bibfield
  {title} {\bibinfo {title} {{3D} stochastic bicontinuous microstructures:
  Generation, topology and elasticity},\ }\href@noop {} {\bibfield  {journal}
  {\bibinfo  {journal} {Acta Materialia}\ }\textbf {\bibinfo {volume} {149}},\
  \bibinfo {pages} {326} (\bibinfo {year} {2018})}\BibitemShut {NoStop}%
\bibitem [{\citenamefont {Aagesen}\ \emph {et~al.}(2010)\citenamefont
  {Aagesen}, \citenamefont {Johnson}, \citenamefont {Fife}, \citenamefont
  {Voorhees}, \citenamefont {Miksis}, \citenamefont {Poulsen}, \citenamefont
  {Lauridsen}, \citenamefont {Marone},\ and\ \citenamefont
  {Stampanoni}}]{Aagesen2010}%
  \BibitemOpen
  \bibfield  {author} {\bibinfo {author} {\bibfnamefont {L.~K.}\ \bibnamefont
  {Aagesen}}, \bibinfo {author} {\bibfnamefont {A.~E.}\ \bibnamefont
  {Johnson}}, \bibinfo {author} {\bibfnamefont {J.~L.}\ \bibnamefont {Fife}},
  \bibinfo {author} {\bibfnamefont {P.~W.}\ \bibnamefont {Voorhees}}, \bibinfo
  {author} {\bibfnamefont {M.~J.}\ \bibnamefont {Miksis}}, \bibinfo {author}
  {\bibfnamefont {S.~O.}\ \bibnamefont {Poulsen}}, \bibinfo {author}
  {\bibfnamefont {E.~M.}\ \bibnamefont {Lauridsen}}, \bibinfo {author}
  {\bibfnamefont {F.}~\bibnamefont {Marone}}, and\ \bibinfo {author}
  {\bibfnamefont {M.}~\bibnamefont {Stampanoni}},\ }\bibfield  {title}
  {\bibinfo {title} {Universality and self-similarity in pinch-off of rods by
  bulk diffusion},\ }\href {https://doi.org/10.1038/nphys1737} {\bibfield
  {journal} {\bibinfo  {journal} {Nature Physics}\ }\textbf {\bibinfo {volume}
  {6}},\ \bibinfo {pages} {796} (\bibinfo {year} {2010})}\BibitemShut {NoStop}%
\bibitem [{\citenamefont {Aagesen}\ \emph {et~al.}(2011)\citenamefont
  {Aagesen}, \citenamefont {Johnson}, \citenamefont {Fife}, \citenamefont
  {Voorhees}, \citenamefont {Miksis}, \citenamefont {Poulsen}, \citenamefont
  {Lauridsen}, \citenamefont {Marone},\ and\ \citenamefont
  {Stampanoni}}]{Aagesen2011}%
  \BibitemOpen
  \bibfield  {author} {\bibinfo {author} {\bibfnamefont {L.}~\bibnamefont
  {Aagesen}}, \bibinfo {author} {\bibfnamefont {A.}~\bibnamefont {Johnson}},
  \bibinfo {author} {\bibfnamefont {J.}~\bibnamefont {Fife}}, \bibinfo {author}
  {\bibfnamefont {P.}~\bibnamefont {Voorhees}}, \bibinfo {author}
  {\bibfnamefont {M.}~\bibnamefont {Miksis}}, \bibinfo {author} {\bibfnamefont
  {S.}~\bibnamefont {Poulsen}}, \bibinfo {author} {\bibfnamefont
  {E.}~\bibnamefont {Lauridsen}}, \bibinfo {author} {\bibfnamefont
  {F.}~\bibnamefont {Marone}}, and\ \bibinfo {author} {\bibfnamefont
  {M.}~\bibnamefont {Stampanoni}},\ }\bibfield  {title} {\bibinfo {title}
  {Pinch-off of rods by bulk diffusion},\ }\href
  {https://doi.org/https://doi.org/10.1016/j.actamat.2011.04.036} {\bibfield
  {journal} {\bibinfo  {journal} {Acta Materialia}\ }\textbf {\bibinfo {volume}
  {59}},\ \bibinfo {pages} {4922 } (\bibinfo {year} {2011})}\BibitemShut
  {NoStop}%
\bibitem [{\citenamefont {Wolterink}\ \emph {et~al.}(2006)\citenamefont
  {Wolterink}, \citenamefont {Barkema},\ and\ \citenamefont
  {Puri}}]{Wolterink2006}%
  \BibitemOpen
  \bibfield  {author} {\bibinfo {author} {\bibfnamefont {J.~K.}\ \bibnamefont
  {Wolterink}}, \bibinfo {author} {\bibfnamefont {G.~T.}\ \bibnamefont
  {Barkema}}, and\ \bibinfo {author} {\bibfnamefont {S.}~\bibnamefont {Puri}},\
  }\bibfield  {title} {\bibinfo {title} {Spinodal decomposition via surface
  diffusion in polymer mixtures},\ }\href
  {https://doi.org/10.1103/PhysRevE.74.011804} {\bibfield  {journal} {\bibinfo
  {journal} {Phys. Rev. E}\ }\textbf {\bibinfo {volume} {74}},\ \bibinfo
  {pages} {011804} (\bibinfo {year} {2006})}\BibitemShut {NoStop}%
\bibitem [{\citenamefont {Takeno}\ and\ \citenamefont
  {Hashimoto}(1998)}]{Takeno1998}%
  \BibitemOpen
  \bibfield  {author} {\bibinfo {author} {\bibfnamefont {H.}~\bibnamefont
  {Takeno}}and\ \bibinfo {author} {\bibfnamefont {T.}~\bibnamefont
  {Hashimoto}},\ }\bibfield  {title} {\bibinfo {title} {Intermittency of
  droplet growth in phase separation of off-critical polymer mixtures},\ }\href
  {https://doi.org/10.1063/1.475484} {\bibfield  {journal} {\bibinfo  {journal}
  {The Journal of Chemical Physics}\ }\textbf {\bibinfo {volume} {108}},\
  \bibinfo {pages} {1225} (\bibinfo {year} {1998})}\BibitemShut {NoStop}%
\bibitem [{\citenamefont {Qian}\ and\ \citenamefont {Chen}(2007)}]{Qian2007}%
  \BibitemOpen
  \bibfield  {author} {\bibinfo {author} {\bibfnamefont {L.~H.}\ \bibnamefont
  {Qian}}and\ \bibinfo {author} {\bibfnamefont {M.~W.}\ \bibnamefont {Chen}},\
  }\bibfield  {title} {\bibinfo {title} {Ultrafine nanoporous gold by
  low-temperature dealloying and kinetics of nanopore formation},\ }\href
  {https://doi.org/10.1063/1.2773757} {\bibfield  {journal} {\bibinfo
  {journal} {Applied Physics Letters}\ }\textbf {\bibinfo {volume} {91}},\
  \bibinfo {pages} {083105} (\bibinfo {year} {2007})}\BibitemShut {NoStop}%
\bibitem [{\citenamefont {Wada}\ \emph {et~al.}(2011)\citenamefont {Wada},
  \citenamefont {Yubuta}, \citenamefont {Inoue},\ and\ \citenamefont
  {Kato}}]{Wada2011}%
  \BibitemOpen
  \bibfield  {author} {\bibinfo {author} {\bibfnamefont {T.}~\bibnamefont
  {Wada}}, \bibinfo {author} {\bibfnamefont {K.}~\bibnamefont {Yubuta}},
  \bibinfo {author} {\bibfnamefont {A.}~\bibnamefont {Inoue}}, and\ \bibinfo
  {author} {\bibfnamefont {H.}~\bibnamefont {Kato}},\ }\bibfield  {title}
  {\bibinfo {title} {Dealloying by metallic melt},\ }\href
  {https://doi.org/https://doi.org/10.1016/j.matlet.2011.01.054} {\bibfield
  {journal} {\bibinfo  {journal} {Materials Letters}\ }\textbf {\bibinfo
  {volume} {65}},\ \bibinfo {pages} {1076 } (\bibinfo {year}
  {2011})}\BibitemShut {NoStop}%
\bibitem [{\citenamefont {Hu}\ \emph {et~al.}(2016)\citenamefont {Hu},
  \citenamefont {Ziehmer}, \citenamefont {Wang},\ and\ \citenamefont
  {Lilleodden}}]{hu2016nanoporous}%
  \BibitemOpen
  \bibfield  {author} {\bibinfo {author} {\bibfnamefont {K.}~\bibnamefont
  {Hu}}, \bibinfo {author} {\bibfnamefont {M.}~\bibnamefont {Ziehmer}},
  \bibinfo {author} {\bibfnamefont {K.}~\bibnamefont {Wang}}, and\ \bibinfo
  {author} {\bibfnamefont {E.~T.}\ \bibnamefont {Lilleodden}},\ }\bibfield
  {title} {\bibinfo {title} {Nanoporous gold: {3D} structural analyses of
  representative volumes and their implications on scaling relations of
  mechanical behaviour},\ }\href@noop {} {\bibfield  {journal} {\bibinfo
  {journal} {Philosophical Magazine}\ }\textbf {\bibinfo {volume} {96}},\
  \bibinfo {pages} {3322} (\bibinfo {year} {2016})}\BibitemShut {NoStop}%
\bibitem [{\citenamefont {Mangipudi}\ \emph {et~al.}(2016)\citenamefont
  {Mangipudi}, \citenamefont {Epler},\ and\ \citenamefont
  {Volkert}}]{mangipudi2016topology}%
  \BibitemOpen
  \bibfield  {author} {\bibinfo {author} {\bibfnamefont {K.}~\bibnamefont
  {Mangipudi}}, \bibinfo {author} {\bibfnamefont {E.}~\bibnamefont {Epler}},
  and\ \bibinfo {author} {\bibfnamefont {C.}~\bibnamefont {Volkert}},\
  }\bibfield  {title} {\bibinfo {title} {Topology-dependent scaling laws for
  the stiffness and strength of nanoporous gold},\ }\href@noop {} {\bibfield
  {journal} {\bibinfo  {journal} {Acta Materialia}\ }\textbf {\bibinfo {volume}
  {119}},\ \bibinfo {pages} {115} (\bibinfo {year} {2016})}\BibitemShut
  {NoStop}%
\bibitem [{\citenamefont {Mullins}(1957)}]{Mullins1957}%
  \BibitemOpen
  \bibfield  {author} {\bibinfo {author} {\bibfnamefont {W.~W.}\ \bibnamefont
  {Mullins}},\ }\bibfield  {title} {\bibinfo {title} {Theory of thermal
  grooving},\ }\href {https://doi.org/10.1063/1.1722742} {\bibfield  {journal}
  {\bibinfo  {journal} {Journal of Applied Physics}\ }\textbf {\bibinfo
  {volume} {28}},\ \bibinfo {pages} {333} (\bibinfo {year} {1957})}\BibitemShut
  {NoStop}%
\bibitem [{\citenamefont {Cahn}\ and\ \citenamefont
  {Hilliard}(1958)}]{CahnHilliard1958}%
  \BibitemOpen
  \bibfield  {author} {\bibinfo {author} {\bibfnamefont {J.~W.}\ \bibnamefont
  {Cahn}}and\ \bibinfo {author} {\bibfnamefont {J.~E.}\ \bibnamefont
  {Hilliard}},\ }\bibfield  {title} {\bibinfo {title} {Free energy of a
  nonuniform system. {I.} interfacial free energy},\ }\href
  {https://doi.org/10.1063/1.1744102} {\bibfield  {journal} {\bibinfo
  {journal} {The Journal of Chemical Physics}\ }\textbf {\bibinfo {volume}
  {28}},\ \bibinfo {pages} {258} (\bibinfo {year} {1958})}\BibitemShut
  {NoStop}%
\bibitem [{\citenamefont {Langer}\ \emph {et~al.}(1975)\citenamefont {Langer},
  \citenamefont {Bar-on},\ and\ \citenamefont {Miller}}]{Langer1975CH}%
  \BibitemOpen
  \bibfield  {author} {\bibinfo {author} {\bibfnamefont {J.~S.}\ \bibnamefont
  {Langer}}, \bibinfo {author} {\bibfnamefont {M.}~\bibnamefont {Bar-on}}, and\
  \bibinfo {author} {\bibfnamefont {H.~D.}\ \bibnamefont {Miller}},\ }\bibfield
   {title} {\bibinfo {title} {New computational method in the theory of
  spinodal decomposition},\ }\href {https://doi.org/10.1103/PhysRevA.11.1417}
  {\bibfield  {journal} {\bibinfo  {journal} {Phys. Rev. A}\ }\textbf {\bibinfo
  {volume} {11}},\ \bibinfo {pages} {1417} (\bibinfo {year}
  {1975})}\BibitemShut {NoStop}%
\bibitem [{\citenamefont {Pego}(1989)}]{Pego1989}%
  \BibitemOpen
  \bibfield  {author} {\bibinfo {author} {\bibfnamefont {R.~L.}\ \bibnamefont
  {Pego}},\ }\bibfield  {title} {\bibinfo {title} {Front migration in the
  nonlinear {Cahn-Hilliard} equation},\ }\href
  {https://doi.org/10.1098/rspa.1989.0027} {\bibfield  {journal} {\bibinfo
  {journal} {Proceedings of the Royal Society of London A: Mathematical,
  Physical and Engineering Sciences}\ }\textbf {\bibinfo {volume} {422}},\
  \bibinfo {pages} {261} (\bibinfo {year} {1989})}\BibitemShut {NoStop}%
\bibitem [{\citenamefont {Cahn}(1961)}]{Cahn1961}%
  \BibitemOpen
  \bibfield  {author} {\bibinfo {author} {\bibfnamefont {J.~W.}\ \bibnamefont
  {Cahn}},\ }\bibfield  {title} {\bibinfo {title} {On spinodal decomposition},\
  }\href {https://doi.org/10.1016/0001-6160(61)90182-1} {\bibfield  {journal}
  {\bibinfo  {journal} {Acta Metallurgica}\ }\textbf {\bibinfo {volume} {9}},\
  \bibinfo {pages} {795} (\bibinfo {year} {1961})}\BibitemShut {NoStop}%
\bibitem [{\citenamefont {Lacasta}\ \emph {et~al.}(1992)\citenamefont
  {Lacasta}, \citenamefont {Hern\'andez-Machado}, \citenamefont {Sancho},\ and\
  \citenamefont {Toral}}]{Lacasta1992}%
  \BibitemOpen
  \bibfield  {author} {\bibinfo {author} {\bibfnamefont {A.~M.}\ \bibnamefont
  {Lacasta}}, \bibinfo {author} {\bibfnamefont {A.}~\bibnamefont
  {Hern\'andez-Machado}}, \bibinfo {author} {\bibfnamefont {J.~M.}\
  \bibnamefont {Sancho}}, and\ \bibinfo {author} {\bibfnamefont
  {R.}~\bibnamefont {Toral}},\ }\bibfield  {title} {\bibinfo {title} {Domain
  growth in binary mixtures at low temperatures},\ }\href
  {https://doi.org/10.1103/PhysRevB.45.5276} {\bibfield  {journal} {\bibinfo
  {journal} {Phys. Rev. B}\ }\textbf {\bibinfo {volume} {45}},\ \bibinfo
  {pages} {5276} (\bibinfo {year} {1992})}\BibitemShut {NoStop}%
\bibitem [{\citenamefont {Zhu}\ \emph {et~al.}(1999)\citenamefont {Zhu},
  \citenamefont {Chen}, \citenamefont {Shen},\ and\ \citenamefont
  {Tikare}}]{Zhu1999}%
  \BibitemOpen
  \bibfield  {author} {\bibinfo {author} {\bibfnamefont {J.}~\bibnamefont
  {Zhu}}, \bibinfo {author} {\bibfnamefont {L.-Q.}\ \bibnamefont {Chen}},
  \bibinfo {author} {\bibfnamefont {J.}~\bibnamefont {Shen}}, and\ \bibinfo
  {author} {\bibfnamefont {V.}~\bibnamefont {Tikare}},\ }\bibfield  {title}
  {\bibinfo {title} {Coarsening kinetics from a variable-mobility
  {Cahn-Hilliard} equation: Application of a semi-implicit {Fourier} spectral
  method},\ }\href {https://doi.org/10.1103/PhysRevE.60.3564} {\bibfield
  {journal} {\bibinfo  {journal} {Phys. Rev. E}\ }\textbf {\bibinfo {volume}
  {60}},\ \bibinfo {pages} {3564} (\bibinfo {year} {1999})}\BibitemShut
  {NoStop}%
\bibitem [{\citenamefont {Cahn}\ \emph {et~al.}(1996)\citenamefont {Cahn},
  \citenamefont {Elliott},\ and\ \citenamefont {Novick-Cohen}}]{Cahn1996sharp}%
  \BibitemOpen
  \bibfield  {author} {\bibinfo {author} {\bibfnamefont {J.~W.}\ \bibnamefont
  {Cahn}}, \bibinfo {author} {\bibfnamefont {C.~M.}\ \bibnamefont {Elliott}},
  and\ \bibinfo {author} {\bibfnamefont {A.}~\bibnamefont {Novick-Cohen}},\
  }\bibfield  {title} {\bibinfo {title} {The {Cahn-Hilliard} equation with a
  concentration dependent mobility: motion by minus the {Laplacian} of the mean
  curvature},\ }\href {https://doi.org/10.1017/S0956792500002369} {\bibfield
  {journal} {\bibinfo  {journal} {European Journal of Applied Mathematics}\
  }\textbf {\bibinfo {volume} {7}},\ \bibinfo {pages} {287} (\bibinfo {year}
  {1996})}\BibitemShut {NoStop}%
\bibitem [{\citenamefont {R\"atz}\ \emph {et~al.}(2006)\citenamefont {R\"atz},
  \citenamefont {Ribalta},\ and\ \citenamefont {Voigt}}]{Ratz2006}%
  \BibitemOpen
  \bibfield  {author} {\bibinfo {author} {\bibfnamefont {A.}~\bibnamefont
  {R\"atz}}, \bibinfo {author} {\bibfnamefont {A.}~\bibnamefont {Ribalta}},
  and\ \bibinfo {author} {\bibfnamefont {A.}~\bibnamefont {Voigt}},\ }\bibfield
   {title} {\bibinfo {title} {Surface evolution of elastically stressed films
  under deposition by a diffuse interface model},\ }\href
  {https://doi.org/10.1016/j.jcp.2005.09.013} {\bibfield  {journal} {\bibinfo
  {journal} {Journal of Computational Physics}\ }\textbf {\bibinfo {volume}
  {214}},\ \bibinfo {pages} {187} (\bibinfo {year} {2006})}\BibitemShut
  {NoStop}%
\bibitem [{\citenamefont {Bray}\ and\ \citenamefont {Emmott}(1995)}]{Bray1995}%
  \BibitemOpen
  \bibfield  {author} {\bibinfo {author} {\bibfnamefont {A.~J.}\ \bibnamefont
  {Bray}}and\ \bibinfo {author} {\bibfnamefont {C.~L.}\ \bibnamefont
  {Emmott}},\ }\bibfield  {title} {\bibinfo {title} {{Lifshitz-Slyozov} scaling
  for late-stage coarsening with an order-parameter-dependent mobility},\
  }\href {https://doi.org/10.1103/PhysRevB.52.R685} {\bibfield  {journal}
  {\bibinfo  {journal} {Physical Review B}\ }\textbf {\bibinfo {volume} {52}},\
  \bibinfo {pages} {R685} (\bibinfo {year} {1995})}\BibitemShut {NoStop}%
\bibitem [{\citenamefont {Dai}\ and\ \citenamefont {Du}(2014)}]{DaiDu2014}%
  \BibitemOpen
  \bibfield  {author} {\bibinfo {author} {\bibfnamefont {S.}~\bibnamefont
  {Dai}}and\ \bibinfo {author} {\bibfnamefont {Q.}~\bibnamefont {Du}},\
  }\bibfield  {title} {\bibinfo {title} {Coarsening mechanism for systems
  governed by the {Cahn-Hilliard} equation with degenerate diffusion
  mobility},\ }\href {https://doi.org/10.1137/140952387} {\bibfield  {journal}
  {\bibinfo  {journal} {Multiscale Modeling \& Simulation}\ }\textbf {\bibinfo
  {volume} {12}},\ \bibinfo {pages} {1870} (\bibinfo {year}
  {2014})}\BibitemShut {NoStop}%
\bibitem [{\citenamefont {Lee}\ \emph {et~al.}(2016)\citenamefont {Lee},
  \citenamefont {M{\"u}nch},\ and\ \citenamefont {S{\"u}li}}]{Lee2016siam}%
  \BibitemOpen
  \bibfield  {author} {\bibinfo {author} {\bibfnamefont {A.}~\bibnamefont
  {Lee}}, \bibinfo {author} {\bibfnamefont {A.}~\bibnamefont {M{\"u}nch}}, and\
  \bibinfo {author} {\bibfnamefont {E.}~\bibnamefont {S{\"u}li}},\ }\bibfield
  {title} {\bibinfo {title} {Sharp-interface limits of the {Cahn-Hilliard}
  equation with degenerate mobility},\ }\href
  {https://doi.org/10.1137/140960189} {\bibfield  {journal} {\bibinfo
  {journal} {SIAM Journal on Applied Mathematics}\ }\textbf {\bibinfo {volume}
  {76}},\ \bibinfo {pages} {433} (\bibinfo {year} {2016})}\BibitemShut
  {NoStop}%
\bibitem [{\citenamefont {Voigt}(2016)}]{Voigt2016comment}%
  \BibitemOpen
  \bibfield  {author} {\bibinfo {author} {\bibfnamefont {A.}~\bibnamefont
  {Voigt}},\ }\bibfield  {title} {\bibinfo {title} {{Comment on "Degenerate
  mobilities in phase field models are insufficient to capture surface
  diffusion" [Appl. Phys. Lett. 107, 081603 (2015)]}},\ }\href
  {https://doi.org/10.1063/1.4939930} {\bibfield  {journal} {\bibinfo
  {journal} {Applied Physics Letters}\ }\textbf {\bibinfo {volume} {108}},\
  \bibinfo {pages} {036101} (\bibinfo {year} {2016})}\BibitemShut {NoStop}%
\bibitem [{\citenamefont {Gugenberger}\ \emph {et~al.}(2008)\citenamefont
  {Gugenberger}, \citenamefont {Spatschek},\ and\ \citenamefont
  {Kassner}}]{Gugenberger2008}%
  \BibitemOpen
  \bibfield  {author} {\bibinfo {author} {\bibfnamefont {C.}~\bibnamefont
  {Gugenberger}}, \bibinfo {author} {\bibfnamefont {R.}~\bibnamefont
  {Spatschek}}, and\ \bibinfo {author} {\bibfnamefont {K.}~\bibnamefont
  {Kassner}},\ }\bibfield  {title} {\bibinfo {title} {Comparison of phase-field
  models for surface diffusion},\ }\href
  {https://doi.org/10.1103/PhysRevE.78.016703} {\bibfield  {journal} {\bibinfo
  {journal} {Phys. Rev. E}\ }\textbf {\bibinfo {volume} {78}},\ \bibinfo
  {pages} {016703} (\bibinfo {year} {2008})}\BibitemShut {NoStop}%
\bibitem [{\citenamefont {Salvalaglio}\ \emph {et~al.}(2019)\citenamefont
  {Salvalaglio}, \citenamefont {Voigt},\ and\ \citenamefont
  {Wise}}]{salvalaglio2019doubly}%
  \BibitemOpen
  \bibfield  {author} {\bibinfo {author} {\bibfnamefont {M.}~\bibnamefont
  {Salvalaglio}}, \bibinfo {author} {\bibfnamefont {A.}~\bibnamefont {Voigt}},
  and\ \bibinfo {author} {\bibfnamefont {S.~M.}\ \bibnamefont {Wise}},\
  }\href@noop {} {\bibinfo {title} {Doubly degenerate diffuse interface models
  of surface diffusion}} (\bibinfo {year} {2019}),\ \Eprint
  {https://arxiv.org/abs/1909.04458} {arXiv:1909.04458 [math.NA]} \BibitemShut
  {NoStop}%
\bibitem [{sup()}]{supplemental}%
  \BibitemOpen
  \href@noop {} {}\bibinfo {note} {See {Supplemental Material} at {[URL]} for
  results and discussion regarding a phase-field model for surface diffusion
  without the stabilizing function present in the {RRV} model. The
  {Supplemental Material} also contains additional data and derivations for the
  {constant-mean-curvature} and {leveled-wave} structures, a description the ad
  hoc convergence criterion we use to define self-similar regimes of evolution
  in our simulations, convergence of the autocorrelations to their self-similar
  states, and details of the conversions between quantities scaled by
  {$S_V^{-1}$} and {$\lambda$}.}\BibitemShut {Stop}%
\bibitem [{\citenamefont {Towns}\ \emph {et~al.}(2014)\citenamefont {Towns},
  \citenamefont {Cockerill}, \citenamefont {Dahan}, \citenamefont {Foster},
  \citenamefont {Gaither}, \citenamefont {Grimshaw}, \citenamefont {Hazlewood},
  \citenamefont {Lathrop}, \citenamefont {Lifka}, \citenamefont {Peterson},
  \citenamefont {Roskies}, \citenamefont {Scott},\ and\ \citenamefont
  {Wilkins-Diehr}}]{xsede}%
  \BibitemOpen
  \bibfield  {author} {\bibinfo {author} {\bibfnamefont {J.}~\bibnamefont
  {Towns}}, \bibinfo {author} {\bibfnamefont {T.}~\bibnamefont {Cockerill}},
  \bibinfo {author} {\bibfnamefont {M.}~\bibnamefont {Dahan}}, \bibinfo
  {author} {\bibfnamefont {I.}~\bibnamefont {Foster}}, \bibinfo {author}
  {\bibfnamefont {K.}~\bibnamefont {Gaither}}, \bibinfo {author} {\bibfnamefont
  {A.}~\bibnamefont {Grimshaw}}, \bibinfo {author} {\bibfnamefont
  {V.}~\bibnamefont {Hazlewood}}, \bibinfo {author} {\bibfnamefont
  {S.}~\bibnamefont {Lathrop}}, \bibinfo {author} {\bibfnamefont
  {D.}~\bibnamefont {Lifka}}, \bibinfo {author} {\bibfnamefont {G.~D.}\
  \bibnamefont {Peterson}}, \bibinfo {author} {\bibfnamefont {R.}~\bibnamefont
  {Roskies}}, \bibinfo {author} {\bibfnamefont {J.}~\bibnamefont {Scott}}, and\
  \bibinfo {author} {\bibfnamefont {N.}~\bibnamefont {Wilkins-Diehr}},\
  }\bibfield  {title} {\bibinfo {title} {{XSEDE}: Accelerating scientific
  discovery},\ }\href {https://doi.org/10.1109/MCSE.2014.80} {\bibfield
  {journal} {\bibinfo  {journal} {Computing in Science \& Engineering}\
  }\textbf {\bibinfo {volume} {16}},\ \bibinfo {pages} {62} (\bibinfo {year}
  {2014})}\BibitemShut {NoStop}%
\bibitem [{\citenamefont {Andrews}\ \emph {et~al.}(2020)\citenamefont
  {Andrews}, \citenamefont {Voorhees},\ and\ \citenamefont
  {Thornton}}]{andrews2020simulation}%
  \BibitemOpen
  \bibfield  {author} {\bibinfo {author} {\bibfnamefont {W.~B.}\ \bibnamefont
  {Andrews}}, \bibinfo {author} {\bibfnamefont {P.~W.}\ \bibnamefont
  {Voorhees}}, and\ \bibinfo {author} {\bibfnamefont {K.}~\bibnamefont
  {Thornton}},\ }\bibfield  {title} {\bibinfo {title} {Simulation of coarsening
  in two-phase systems with dissimilar mobilities},\ }\href@noop {} {\bibfield
  {journal} {\bibinfo  {journal} {Computational Materials Science}\ }\textbf
  {\bibinfo {volume} {173}},\ \bibinfo {pages} {109418} (\bibinfo {year}
  {2020})}\BibitemShut {NoStop}%
\bibitem [{\citenamefont {Marsh}\ and\ \citenamefont
  {Glicksman}(1996)}]{Marsh1996}%
  \BibitemOpen
  \bibfield  {author} {\bibinfo {author} {\bibfnamefont {S.~P.}\ \bibnamefont
  {Marsh}}and\ \bibinfo {author} {\bibfnamefont {M.~E.}\ \bibnamefont
  {Glicksman}},\ }\bibfield  {title} {\bibinfo {title} {Overview of geometric
  effects on coarsening of mushy zones},\ }\href
  {https://doi.org/10.1007/BF02648946} {\bibfield  {journal} {\bibinfo
  {journal} {Metallurgical and Materials Transactions A}\ }\textbf {\bibinfo
  {volume} {27}},\ \bibinfo {pages} {557} (\bibinfo {year} {1996})}\BibitemShut
  {NoStop}%
\bibitem [{\citenamefont {Park}\ \emph {et~al.}(2014)\citenamefont {Park},
  \citenamefont {Voorhees},\ and\ \citenamefont {Thornton}}]{Park2014}%
  \BibitemOpen
  \bibfield  {author} {\bibinfo {author} {\bibfnamefont {C.-L.}\ \bibnamefont
  {Park}}, \bibinfo {author} {\bibfnamefont {P.~W.}\ \bibnamefont {Voorhees}},
  and\ \bibinfo {author} {\bibfnamefont {K.}~\bibnamefont {Thornton}},\
  }\bibfield  {title} {\bibinfo {title} {Application of the level-set method to
  the analysis of an evolving microstructure},\ }\href
  {https://doi.org/10.1016/j.commatsci.2013.12.022} {\bibfield  {journal}
  {\bibinfo  {journal} {Computational Materials Science}\ }\textbf {\bibinfo
  {volume} {85}},\ \bibinfo {pages} {46} (\bibinfo {year} {2014})}\BibitemShut
  {NoStop}%
\bibitem [{\citenamefont {Odgaard}\ and\ \citenamefont
  {Gundersen}(1993)}]{Odgaard1993}%
  \BibitemOpen
  \bibfield  {author} {\bibinfo {author} {\bibfnamefont {A.}~\bibnamefont
  {Odgaard}}and\ \bibinfo {author} {\bibfnamefont {H.}~\bibnamefont
  {Gundersen}},\ }\bibfield  {title} {\bibinfo {title} {Quantification of
  connectivity in cancellous bone, with special emphasis on {3-D}
  reconstructions},\ }\href
  {https://doi.org/https://doi.org/10.1016/8756-3282(93)90245-6} {\bibfield
  {journal} {\bibinfo  {journal} {Bone}\ }\textbf {\bibinfo {volume} {14}},\
  \bibinfo {pages} {173 } (\bibinfo {year} {1993})}\BibitemShut {NoStop}%
\bibitem [{\citenamefont {Weisstein}()}]{MWEuler}%
  \BibitemOpen
  \bibfield  {author} {\bibinfo {author} {\bibfnamefont {E.~W.}\ \bibnamefont
  {Weisstein}},\ }\href {http://mathworld.wolfram.com/EulerCharacteristic.html}
  {\bibinfo {title} {Euler characteristic. {From MathWorld---A Wolfram Web
  Resource}}},\ \bibinfo {note} {last visited on 11/21/2018}\BibitemShut
  {NoStop}%
\bibitem [{\citenamefont {Sun}\ \emph {et~al.}(2017)\citenamefont {Sun},
  \citenamefont {Cecen}, \citenamefont {Gibbs}, \citenamefont {Kalidindi},\
  and\ \citenamefont {Voorhees}}]{Sun2017}%
  \BibitemOpen
  \bibfield  {author} {\bibinfo {author} {\bibfnamefont {Y.}~\bibnamefont
  {Sun}}, \bibinfo {author} {\bibfnamefont {A.}~\bibnamefont {Cecen}}, \bibinfo
  {author} {\bibfnamefont {J.}~\bibnamefont {Gibbs}}, \bibinfo {author}
  {\bibfnamefont {S.}~\bibnamefont {Kalidindi}}, and\ \bibinfo {author}
  {\bibfnamefont {P.}~\bibnamefont {Voorhees}},\ }\bibfield  {title} {\bibinfo
  {title} {Analytics on large micsotructure datasets using two-point spatial
  correlations: Coarsening of dendritic structure},\ }\href
  {https://doi.org/10.1016/j.actamat.2017.04.054} {\bibfield  {journal}
  {\bibinfo  {journal} {Acta Materialia}\ }\textbf {\bibinfo {volume} {132}},\
  \bibinfo {pages} {374} (\bibinfo {year} {2017})}\BibitemShut {NoStop}%
\bibitem [{\citenamefont {Sun}(2018)}]{Sun2018}%
  \BibitemOpen
  \bibfield  {author} {\bibinfo {author} {\bibfnamefont {Y.}~\bibnamefont
  {Sun}},\ }\emph {\bibinfo {title} {Spatio-Temporal Analysis of Coarsening in
  Complex Microstructures Using Two-Point Statistics}},\ \href@noop {} {Ph.D.
  thesis},\ \bibinfo  {school} {Northwestern University} (\bibinfo {year}
  {2018})\BibitemShut {NoStop}%
\bibitem [{\citenamefont {Sun}\ \emph {et~al.}(2019)\citenamefont {Sun},
  \citenamefont {Elder},\ and\ \citenamefont {Voorhees}}]{Sun_2019}%
  \BibitemOpen
  \bibfield  {author} {\bibinfo {author} {\bibfnamefont {Y.}~\bibnamefont
  {Sun}}, \bibinfo {author} {\bibfnamefont {K.~L.~M.}\ \bibnamefont {Elder}},
  and\ \bibinfo {author} {\bibfnamefont {P.~W.}\ \bibnamefont {Voorhees}},\
  }\bibfield  {title} {\bibinfo {title} {Morphological characterization of
  bicontinuous microstructures using two-point statistics},\ }\href
  {https://doi.org/10.1088/1757-899x/580/1/012011} {\bibfield  {journal}
  {\bibinfo  {journal} {{IOP} Conference Series: Materials Science and
  Engineering}\ }\textbf {\bibinfo {volume} {580}},\ \bibinfo {pages} {012011}
  (\bibinfo {year} {2019})}\BibitemShut {NoStop}%
\bibitem [{\citenamefont {Debye}\ \emph {et~al.}(1957)\citenamefont {Debye},
  \citenamefont {Anderson~Jr},\ and\ \citenamefont
  {Brumberger}}]{debye1957scattering}%
  \BibitemOpen
  \bibfield  {author} {\bibinfo {author} {\bibfnamefont {P.}~\bibnamefont
  {Debye}}, \bibinfo {author} {\bibfnamefont {H.}~\bibnamefont {Anderson~Jr}},
  and\ \bibinfo {author} {\bibfnamefont {H.}~\bibnamefont {Brumberger}},\
  }\bibfield  {title} {\bibinfo {title} {Scattering by an inhomogeneous solid.
  {II}. the correlation function and its application},\ }\href@noop {}
  {\bibfield  {journal} {\bibinfo  {journal} {Journal of Applied Physics}\
  }\textbf {\bibinfo {volume} {28}},\ \bibinfo {pages} {679} (\bibinfo {year}
  {1957})}\BibitemShut {NoStop}%
\bibitem [{\citenamefont {Teubner}(1990)}]{Teubner1990}%
  \BibitemOpen
  \bibfield  {author} {\bibinfo {author} {\bibfnamefont {M.}~\bibnamefont
  {Teubner}},\ }\bibfield  {title} {\bibinfo {title} {Scattering from two-phase
  random media},\ }\href@noop {} {\bibfield  {journal} {\bibinfo  {journal}
  {Journal of Chemical Physics}\ }\textbf {\bibinfo {volume} {92}},\ \bibinfo
  {pages} {4501} (\bibinfo {year} {1990})}\BibitemShut {NoStop}%
\bibitem [{\citenamefont {Porod}(1951)}]{porod1951}%
  \BibitemOpen
  \bibfield  {author} {\bibinfo {author} {\bibfnamefont {G.}~\bibnamefont
  {Porod}},\ }\bibfield  {title} {\bibinfo {title} {Die
  {R\"ontgenkleinwinkelstreuung} von dichtgepackten kolloiden {Systemen}},\
  }\href@noop {} {\bibfield  {journal} {\bibinfo  {journal} {Kolloid-Z.}\
  }\textbf {\bibinfo {volume} {124}},\ \bibinfo {pages} {83} (\bibinfo {year}
  {1951})}\BibitemShut {NoStop}%
\bibitem [{\citenamefont {Bouttes}\ \emph {et~al.}(2016)\citenamefont
  {Bouttes}, \citenamefont {Gouillart},\ and\ \citenamefont
  {Vandembroucq}}]{Bouttes2016}%
  \BibitemOpen
  \bibfield  {author} {\bibinfo {author} {\bibfnamefont {D.}~\bibnamefont
  {Bouttes}}, \bibinfo {author} {\bibfnamefont {E.}~\bibnamefont {Gouillart}},
  and\ \bibinfo {author} {\bibfnamefont {D.}~\bibnamefont {Vandembroucq}},\
  }\bibfield  {title} {\bibinfo {title} {Topological symmetry breaking in
  viscous coarsening},\ }\href {https://doi.org/10.1103/PhysRevLett.117.145702}
  {\bibfield  {journal} {\bibinfo  {journal} {Phys. Rev. Lett.}\ }\textbf
  {\bibinfo {volume} {117}},\ \bibinfo {pages} {145702} (\bibinfo {year}
  {2016})}\BibitemShut {NoStop}%
\bibitem [{\citenamefont {Bernoff}\ \emph {et~al.}(1998)\citenamefont
  {Bernoff}, \citenamefont {Bertozzi},\ and\ \citenamefont
  {Witelski}}]{Bernoff1998}%
  \BibitemOpen
  \bibfield  {author} {\bibinfo {author} {\bibfnamefont {A.~J.}\ \bibnamefont
  {Bernoff}}, \bibinfo {author} {\bibfnamefont {A.~L.}\ \bibnamefont
  {Bertozzi}}, and\ \bibinfo {author} {\bibfnamefont {T.~P.}\ \bibnamefont
  {Witelski}},\ }\bibfield  {title} {\bibinfo {title} {Axisymmetric surface
  diffusion: Dynamics and stability of self-similar pinchoff},\ }\href
  {https://doi.org/10.1023/B:JOSS.0000033251.81126.af} {\bibfield  {journal}
  {\bibinfo  {journal} {Journal of Statistical Physics}\ }\textbf {\bibinfo
  {volume} {93}},\ \bibinfo {pages} {725} (\bibinfo {year} {1998})}\BibitemShut
  {NoStop}%
\bibitem [{\citenamefont {Wong}\ \emph {et~al.}(1998)\citenamefont {Wong},
  \citenamefont {Miksis}, \citenamefont {Voorhees},\ and\ \citenamefont
  {Davis}}]{Wong1998}%
  \BibitemOpen
  \bibfield  {author} {\bibinfo {author} {\bibfnamefont {H.}~\bibnamefont
  {Wong}}, \bibinfo {author} {\bibfnamefont {M.~J.}\ \bibnamefont {Miksis}},
  \bibinfo {author} {\bibfnamefont {P.~W.}\ \bibnamefont {Voorhees}}, and\
  \bibinfo {author} {\bibfnamefont {S.~H.}\ \bibnamefont {Davis}},\ }\bibfield
  {title} {\bibinfo {title} {Universal pinch off of rods by capillarity-driven
  surface diffusion},\ }\bibfield  {journal} {\bibinfo  {journal} {Scripta
  Materialia}\ }\textbf {\bibinfo {volume} {39}},\ \href
  {https://doi.org/10.1016/S1359-6462(98)00127-4}
  {10.1016/S1359-6462(98)00127-4} (\bibinfo {year} {1998})\BibitemShut
  {NoStop}%
\bibitem [{\citenamefont {Gorshkov}\ and\ \citenamefont
  {Privman}(2017)}]{Gorshkov2017}%
  \BibitemOpen
  \bibfield  {author} {\bibinfo {author} {\bibfnamefont {V.}~\bibnamefont
  {Gorshkov}}and\ \bibinfo {author} {\bibfnamefont {V.}~\bibnamefont
  {Privman}},\ }\bibfield  {title} {\bibinfo {title} {Kinetic {Monte} {Carlo}
  model of breakup of nanowires into chains of nanoparticles},\ }\href
  {https://doi.org/10.1063/1.5002665} {\bibfield  {journal} {\bibinfo
  {journal} {Journal of Applied Physics}\ }\textbf {\bibinfo {volume} {122}},\
  \bibinfo {pages} {204301} (\bibinfo {year} {2017})}\BibitemShut {NoStop}%
\bibitem [{\citenamefont {Lai}\ \emph {et~al.}(2019)\citenamefont {Lai},
  \citenamefont {Han}, \citenamefont {Spurgeon}, \citenamefont {Huang},
  \citenamefont {Thiel}, \citenamefont {Liu},\ and\ \citenamefont
  {Evans}}]{Lai2019}%
  \BibitemOpen
  \bibfield  {author} {\bibinfo {author} {\bibfnamefont {K.~C.}\ \bibnamefont
  {Lai}}, \bibinfo {author} {\bibfnamefont {Y.}~\bibnamefont {Han}}, \bibinfo
  {author} {\bibfnamefont {P.}~\bibnamefont {Spurgeon}}, \bibinfo {author}
  {\bibfnamefont {W.}~\bibnamefont {Huang}}, \bibinfo {author} {\bibfnamefont
  {P.~A.}\ \bibnamefont {Thiel}}, \bibinfo {author} {\bibfnamefont {D.-J.}\
  \bibnamefont {Liu}}, and\ \bibinfo {author} {\bibfnamefont {J.~W.}\
  \bibnamefont {Evans}},\ }\bibfield  {title} {\bibinfo {title} {Reshaping,
  intermixing, and coarsening for metallic nanocrystals: Nonequilibrium
  statistical mechanical and coarse-grained modeling},\ }\href
  {https://doi.org/10.1021/acs.chemrev.8b00582} {\bibfield  {journal} {\bibinfo
   {journal} {Chemical Reviews}\ }\textbf {\bibinfo {volume} {119}},\ \bibinfo
  {pages} {6670} (\bibinfo {year} {2019})}\BibitemShut {NoStop}%
\bibitem [{\citenamefont {Fife}\ \emph {et~al.}(2014)\citenamefont {Fife},
  \citenamefont {Gibbs}, \citenamefont {Gulsoy}, \citenamefont {Park},
  \citenamefont {Thornton},\ and\ \citenamefont {Voorhees}}]{Fife2014}%
  \BibitemOpen
  \bibfield  {author} {\bibinfo {author} {\bibfnamefont {J.~L.}\ \bibnamefont
  {Fife}}, \bibinfo {author} {\bibfnamefont {J.~W.}\ \bibnamefont {Gibbs}},
  \bibinfo {author} {\bibfnamefont {E.~B.}\ \bibnamefont {Gulsoy}}, \bibinfo
  {author} {\bibfnamefont {C.-L.}\ \bibnamefont {Park}}, \bibinfo {author}
  {\bibfnamefont {K.}~\bibnamefont {Thornton}}, and\ \bibinfo {author}
  {\bibfnamefont {P.~W.}\ \bibnamefont {Voorhees}},\ }\bibfield  {title}
  {\bibinfo {title} {The dynamics of interfaces during coarsening in
  solid-liquid systems},\ }\href
  {https://doi.org/10.1016/j.actamat.2014.01.024} {\bibfield  {journal}
  {\bibinfo  {journal} {Acta Materialia}\ }\textbf {\bibinfo {volume} {70}},\
  \bibinfo {pages} {66} (\bibinfo {year} {2014})}\BibitemShut {NoStop}%
\end{thebibliography}
\end{document}